\documentclass[prd,aps,a4paper,nofootinbib,twocolumn]{revtex4-1}  

\newif\ifusesec
\usesectrue  
   
\usepackage{graphicx} 
\usepackage{mathrsfs}
\usepackage{amsmath,amsfonts,amssymb}
\usepackage{multirow}

\newcommand{\beq}{\begin{equation}}
\newcommand{\eeq}{\end{equation}}
\begin{document}

\title{New gravitational self-force analytical results for \\ eccentric orbits around a Schwarzschild black hole}

\author{Donato \surname{Bini}$^1$}
\author{Thibault \surname{Damour}$^2$}
\author{Andrea \surname{Geralico}$^1$}

\affiliation{$^1$Istituto per le Applicazioni del Calcolo ``M. Picone'', CNR, I-00185 Rome, Italy\\
$^2$Institut des Hautes Etudes Scientifiques, 91440 Bures-sur-Yvette, France}

\date{\today}

\begin{abstract}
We raise the analytical knowledge of the eccentricity-expansion of the Detweiler-Barack-Sago redshift invariant in a Schwarzschild spacetime 
 up to the 9.5th post-Newtonian order (included)  for the $e^2$ and $e^4$ contributions, and up to the 4th post-Newtonian order for the higher eccentricity contributions
 through $e^{20}$.  We convert this information into an analytical knowledge of the effective-one-body radial potentials $\bar d(u)$, $\rho(u)$ and $q(u)$ through the 9.5th post-Newtonian order. We find that our analytical results are compatible with
 current corresponding numerical self-force data.
\end{abstract}

\maketitle

\section{Introduction}

A fruitful {\it synergy} between  various methods for  approximating the general relativistic two-body problem has developed over
the last years, with accelerated progress over the last months.  The concerned approximation methods are: post-Newtonian (PN)
theory, self-force (SF) theory, and numerical relativity (NR). The synergy between these approximation methods was greatly
facilitated by the construction of  theoretical  {\it bridges} connecting the various methods.  Among these bridges, two have been
particularly useful: the effective-one-body (EOB) formalism \cite{Buonanno:1998gg,Buonanno:2000ef,Damour:2000we,Damour:2001tu},
and the first law of binary mechanics \cite{LeTiec:2011ab,Blanchet:2012at,Tiec:2015cxa}.  Examples of synergies between PN and SF
facilitated by EOB and/or the first law are Refs. \cite{Damour:2009sm,Barack:2010ny,LeTiec:2011dp,Barausse:2011dq,Akcay:2012ea,Shah:2013uya,%Bini:2014nfa,
Dolan:2013roa,Keidl:2010pm,Shah:2012gu,Bini:2013zaa,Bini:2013rfa,Bini:2014ica,Dolan:2014pja,Bini:2014zxa,Bini:2015bla,Bini:2015mza,Akcay:2015pza,Kavanagh:2015lva,vandeMeent:2015lxa,Bini:2015bfb,Hopper:2015icj,Akcay:2015pjz,Shah:2015nva}.

This paper is a follow-up of Ref. \cite{Bini:2015bfb}. It concerns  the  first self-force (1SF) conservative dynamics of the eccentric orbits of a small mass $m_1$ around a
(non-spinning) large mass $m_2$ (described by a Schwarzschild black hole). Our results complete the results of both Ref. \cite{Bini:2015bfb} and of the recent related 
Refs. \cite{Hopper:2015icj,Akcay:2015pjz}. Before entering the details of our new results
we  summarize in Table \ref{tab:1}  how our results go beyond present analytical knowledge in terms of the decomposition of the 
(gauge-invariant) 1SF contribution $\delta U(p,e)$ to the  Detweiler-Barack-Sago \cite{Detweiler:2008ft,Barack:2011ed} average redshift $U(p,e)$ in powers of the eccentricity $e$, i.e., $\delta U(p,e)=\sum_n \delta U^{e^n}(u_p) e^n$. 
[We use in the present paper the same notation as in Ref. \cite{Bini:2015bfb}.  
In particular, $u_p \equiv 1/p$ denotes the inverse semi-latus rectum of the considered eccentric orbit.
In addition, we denote $M\equiv m_1+m_2$, $\mu\equiv m_1m_2/(m_1+m_2)$ and $\nu\equiv \mu/M=m_1m_2/(m_1+m_2)^2$ in our EOB considerations.]

\begin{table}
\centering
\caption{Present analytical knowledge of  $\delta U^{e^n}$ along eccentric orbits in a Schwarzschild spacetime.}
\begin{ruledtabular}
\begin{tabular}{ccc}
$n$ & $\delta U^{e^n}$ & Refs.\cr
\hline
0   &22.5PN&  Kavanagh et al.\cite{Kavanagh:2015lva}\cr  
2   &9.5PN &  This paper\cr
4   &9.5PN&  This paper\cr 
6   &4PN&  Hopper et al.\cite{Hopper:2015icj}\cr 
8   &4PN&  Hopper et al.\cite{Hopper:2015icj}\cr 
10   &4PN&  Hopper et al.\cite{Hopper:2015icj} \cr 
12  &4PN &  This paper \cr 
14  &4PN &   This paper \cr 
16  &4PN &  This paper \cr 
18  &4PN &  This paper \cr 
20  &4PN &   This paper \cr 
\end{tabular}
\end{ruledtabular}
\label{tab:1}
\end{table}

Table \ref{tab:1} shows that our new results are of two different types. On the one hand, we improve the PN knowledge of the contributions to $\delta U$ of order $e^2$ and $e^4$ to the 9.5PN level (previous analytical knowledge was the 6.5PN level for $\delta U^{e^2}$ \cite{Bini:2015bfb} and the 4PN one for $\delta U^{e^4}$  \cite{Damour:2015isa,Bini:2015bfb}). 
On the other hand, we combine the 4PN results of    \cite{Damour:2015isa} with the eccentric first law \cite{Tiec:2015cxa} to compute the 4PN-accurate values of  $\delta U^{e^n}$
for the high values of $n$: $n=12, 14, 16, 18, 20$ (previous similar 4PN-level knowledge concerned $n= 6, 8, 10$ \cite{Hopper:2015icj}). [We also give below the 4PN knowledge of
the corresponding high eccentricity powers of  the alternative redshift function
$\delta z_1(p,e)=\sum_n \delta z_1^{e^n}(u_p) e^n$, where $z_1=1/U$.]

To complete our results on the coefficients at orders $e^2$ and $e^4$ of the redshift function $\delta U(p,e)=\sum_n \delta U^{e^n}(u_p) e^n$,  we shall also transcribe below our 
9.5PN-accurate results 
in terms of the corresponding EOB potentials $\bar d(u)$ and $q(u) \equiv q_4(u)$. [We also give the previously uncomputed 4PN values of the higher-$p_r$-powers analogs
of the $O(p_r^4)$ EOB potential   $q(u) \equiv q_4(u)$.] 

Finally, we shall also explicitly compute the 9.5PN-accurate value of the gauge-invariant 1SF precession function $\rho(u)$  defined in Ref. \cite{Damour:2009sm} and related there
to the 1SF  EOB potentials $a(u)$ and $\bar d(u)$. The precession function  $\rho(u)$ is of particular interest because it can be directly extracted from SF numerical computations 
of the dynamics of slightly eccentric orbits \cite{Barack:2010ny} {\it without} making use of the eccentric first law. 
Therefore a comparison between our 9.5PN analytical 
computation of the precession function $\rho(u)$ (which combines SF theory with the eccentric first law \cite{Tiec:2015cxa}) and of a purely dynamical SF numerical computation of the precession of eccentric orbits (as in \cite{Barack:2010ny}) would be a useful check of the assumptions underlying the theoretical bridges (EOB and the first law) used in connecting SF and PN results.

\section{Novel analytical results for $\delta U^{e^2}$ and $\delta U^{e^4}$ up to the 9.5PN order}

Our new, 9.5PN-accurate, results for $\delta U^{e^2}$ and $\delta U^{e^4}$ have been obtained by following the approach of our previous papers
 \cite{Bini:2013zaa,Bini:2015bfb}. Let us only recall that our approach combines standard Regge-Wheeler-Zerilli first order perturbation theory with the
 Mano-Suzuki-Takasugi (MST) \cite{Mano:1996mf,Mano:1996vt} hypergeometric-expansion technique (here used up to the multipolar $l=7$ solution included).
The main steps of this, by now, well established procedure are sketched in Appendix \ref{appRWZ}.

We have raised the analytical knowledge of $\delta U^{e^2}$ from the 6.5PN level obtained in our previous work  \cite{Bini:2015bfb}  up to the 9.5PN level.
Note that the conversion between PN order and meaningful powers of $u_p$,  or equivalently\footnote{The name we give to the arguments in the various EOB potentials
considered here is arbitrary, because we are expanding the corresponding {\it functions} (e.g. $u \to \bar d(u)$) in powers of their argument. 
The traditional EOB notation for the argument is $u$, but as the corresponding
physical quantity $u= GM/c^2 r_{\rm EOB}$ is numerically equal, modulo a $O(\nu)$ correction,  both to $u_p=1/p$ and to the frequency parameter usually denoted $x$, one
sometimes calls the argument $u_p$ or $x$.}   $u$ or $x$,  depends on the considered SF or EOB function. More precisely,  the $n$th PN order corresponds to:
(1) a term $\propto  u^{n+1}$ in $a(u)$ (and $\delta U(u,e)$); (2)    a term $\propto  u^n$ in $\bar d (u)$ or $\rho(u)$; and  (3)  a term $\propto  u^{n-1}$ in $q(u)$.  Therefore, our current 9.5PN accuracy (obtained by using hypergeometric expansions up to the multipolar order $l=7$) corresponds to error terms: $O_{\ln{}}(u_p^{11})$ in  $\delta U(u_p,e)$;  $O_{\ln{}}(u^{10})$ in  $\bar d (u)$ or $\rho(u)$; and $O_{\ln{}}(u^{9})$ in  $ q (u)$, where $O_{\ln{}}(u^{q})$ denotes some $O(u^q (\ln u)^p)$ with a non specified natural integer $p\ge 1$. Our result for $\delta U^{e^2}$ reads

\begin{widetext}

\begin{eqnarray}
\label{DeltaU_e2}
\delta U^{e^2}(u_p) &=&  u_p+4 u_p^2+7 u_p^3+\left(-\frac{5}{3}-\frac{41}{32}\pi^2\right) u_p^4\nonumber\\
&&+\left(-\frac{11141}{45}+\frac{29665}{3072}\pi^2-\frac{296}{15}\ln(u_p)-\frac{592}{15}\gamma-\frac{1458}{5}\ln(3)+\frac{3248}{15}\ln(2)\right) u_p^5\nonumber\\
&&+\left(-\frac{2238629}{1575}-\frac{73145}{1536}\pi^2+\frac{8696}{105}\ln(u_p)-\frac{167696}{105}\ln(2)+\frac{17392}{105}\gamma+\frac{42282}{35}\ln(3)\right) u_p^6\nonumber\\
&&-\frac{232618}{1575}\pi u_p^{13/2}\nonumber\\
&& +\left(\frac{2750367763}{198450}-\frac{9765625}{4536}\ln(5)+\frac{41285072}{2835}\ln(2)+\frac{5102288}{2835}\gamma-\frac{673353}{280}\ln(3)+\frac{2551144}{2835}\ln(u_p)\right. \nonumber\\
&& \left. +\frac{9735101}{262144}\pi^4-\frac{13433142863}{3538944}\pi^2\right) u_p^7\nonumber\\
&&+\frac{2687231}{4410}\pi u_p^{15/2}\nonumber\\
&&+\left[\frac{1040896}{1575}\ln(u_p)^2+\left(\frac{4163584}{1575}\gamma-\frac{85422206699}{5457375}+\frac{936036}{175}\ln(3)-\frac{109568}{1575}\ln(2)\right)\ln(u_p)
\right. \nonumber\\
&& +\frac{471677766820151}{1719073125}-\frac{171448137814}{5457375}\ln(2)-\frac{301990638447}{4312000}\ln(3)+\frac{1228515625}{57024}\ln(5)-\frac{170844413398}{5457375}\gamma\nonumber\\
&&+\frac{1872072}{175}\gamma\ln(3)-\frac{219136}{1575}\ln(2)\gamma+\frac{1872072}{175}\ln(2)\ln(3)+\frac{4163584}{1575}\gamma^2-\frac{23854914937}{503316480}\pi^4\nonumber\\
&& \left. +\frac{936036}{175}\ln(3)^2-\frac{77824}{15}\zeta(3)-\frac{8655872}{1575}\ln(2)^2-\frac{80420758955297}{2477260800}\pi^2\right] u_p^8\nonumber\\
&&+\frac{66757650913}{26195400}\pi u_p^{17/2}\nonumber
\end{eqnarray}
\begin{eqnarray}
\label{DeltaU_e2}
\phantom{\delta U^{e^2}(u_p)} 
&&+\left[-\frac{2994904}{1225}\ln(u_p)^2+\left(-\frac{11979616}{1225}\gamma+\frac{618506181077}{99324225}-\frac{5165694}{175}\ln(3)+\frac{55690528}{2205}\ln(2)\right)\ln(u_p)\right.\nonumber\\
&&-\frac{2205806334400049687}{1720792198125}-\frac{46585620571706}{165540375}\ln(2)+\frac{2452189382919}{8008000}\ln(3)-\frac{3191857421875}{36324288}\ln(5)\nonumber\\
&&-\frac{678223072849}{46332000}\ln(7)+\frac{1272610164394}{99324225}\gamma-\frac{10331388}{175}\gamma\ln(3)+\frac{111381056}{2205}\ln(2)\gamma\nonumber\\
&& -\frac{10331388}{175}\ln(2)\ln(3)-\frac{11979616}{1225}\gamma^2+\frac{389897083139633}{16106127360}\pi^4\nonumber\\
&& \left. -\frac{5165694}{175}\ln(3)^2+\frac{1020736}{105}\zeta(3)+\frac{1391778208}{11025}\ln(2)^2-\frac{79965804866374541}{554906419200}\pi^2\right] u_p^9\nonumber\\
&&
+\left(-\frac{3936830890988503}{59935075200}\pi+\frac{100155852}{6125}\pi\ln(3)-\frac{2250424}{675}\pi^3+\frac{120397684}{23625}\pi\ln(u_p)+\frac{665599064}{165375}\pi\ln(2)\right. \nonumber\\
&& \left. +\frac{240795368}{23625}\pi\gamma\right) u_p^{19/2}\nonumber\\
&&
+\left[-\frac{91608512384}{9823275}\ln(u_p)^2+\left(\frac{2694566979}{53900}\ln(3)+\frac{105972007312260412}{442489422375}+\frac{76708984375}{1571724}\ln(5)\right.\right.\nonumber\\
&&\left.
-\frac{366434049536}{9823275}\gamma-\frac{2995825170944}{9823275}\ln(2)\right)\ln(u_p) 
-\frac{86555681446617433123159}{949139810994375}+\frac{2694566979}{53900}\ln(3)^2\nonumber\\
&&
+\frac{213354316911514424}{442489422375}\gamma+\frac{1656928811171577752}{442489422375}\ln(2)-\frac{995870224363383}{1079078000}\ln(3)-\frac{38345561821484375}{56638646064}\ln(5)\nonumber\\
&&
+\frac{315073184}{2835}\zeta(3)+\frac{193778020814}{868725}\ln(7)-\frac{113425393373}{100663296}\pi^6-\frac{3608718872135173}{5651824640}\pi^2\nonumber\\
&&
+\frac{16005605256259137079}{16492674416640}\pi^4+\frac{76708984375}{1571724}\ln(5)^2-\frac{366434049536}{9823275}\gamma^2-\frac{5991650341888}{9823275}\ln(2)\gamma\nonumber\\
&& 
+\frac{76708984375}{785862}\ln(2)\ln(5)-\frac{239758989824}{218295}\ln(2)^2+\frac{2694566979}{26950}\gamma\ln(3)+\frac{76708984375}{785862}\gamma\ln(5)\nonumber\\
&&\left.
+\frac{2694566979}{26950}\ln(2)\ln(3)\right]u_p^{10}\nonumber\\
&& 
+\left(-\frac{28108289357}{1157625}\pi\ln(u_p)+\frac{369663722}{33075}\pi^3-\frac{56216578714}{1157625}\pi\gamma+\frac{15720247936467024947}{114535928707200}\pi\right.\nonumber\\
&& \left.
+\frac{9003848366}{231525}\pi\ln(2)-\frac{839692089}{8575}\pi\ln(3) \right)u_p^{21/2}
+O_{\ln{}}(u_p^{11})\,.
\end{eqnarray}

The numerical values of the coefficients in the latter expansion read
\begin{eqnarray}
\delta U^{e^2}(u_p) &=&u_p+4 u_p^2+7 u_p^3-14.31209731 u_p^4+(-345.3178497-19.73333333\ln(u_p)) u_p^5\nonumber\\
&& +(-1575.580014+82.81904762\ln(u_p)) u_p^6-463.9942859 u_p^{13/2}\nonumber\\
&&+(-14960.48992+899.8744268\ln(u_p)) u_p^7+1914.327703 u_p^{15/2}\nonumber\\
&& +(-119420.1688-8298.710150\ln(u_p)+660.8863492\ln(u_p)^2) u_p^8+8006.189854 u_p^{17/2}\nonumber\\
&& +(-395945.586-14340.26852\ln(u_p)-2444.819592\ln(u_p)^2) u_p^9\nonumber\\
&& +(-226044.9538+16010.17903 \ln(u_p)) u_p^{19/2}\nonumber\\
&& +(140039.6684\ln(u_p)-9325.658946\ln(u_p)^2-2966833.394) u_p^{10}\nonumber\\
&& +(-76281.00237\ln(u_p)+436383.4353) u_p^{21/2}
+O_{\ln{}}(u_p^{11})\,.
\end{eqnarray}

Similarly, we have extended the analytical knowledge of $\delta U^{e^4}$ from 4PN (as obtained in our previous work  \cite{Bini:2015bfb}) up to 9.5PN, namely
\begin{eqnarray}
\label{DeltaU_e4}
\delta U^{e^4}(u_p) &=& -2 u_p^2+\frac14 u_p^3+\left(\frac{705}{8}-\frac{123}{256}\pi^2\right) u_p^4\nonumber\\
&+& \left(\frac{247931}{360}-\frac{89395}{6144}\pi^2+\frac{28431}{10}\ln(3)+\frac{292}{3}\gamma-\frac{64652}{15}\ln(2)+\frac{146}{3}\ln(u_p)\right)u_p^5\nonumber\\
&+& \left(\frac{293423}{4200} -\frac{25493859}{2240}\ln(3)-\frac{601}{5}\ln(u_p)-\frac{1202}{5}\gamma+\frac{248378}{7}\ln(2)-\frac{9765625}{1344}\ln(5)+\frac{275167}{1024}\pi^2\right)u_p^6\nonumber\\
&+&  \frac{430889}{3150}\pi u_p^{13/2}\nonumber\\
&&+\left(-\frac{4815135047}{396900}-\frac{194385796}{945}\ln(2)-\frac{2260629}{320}\ln(3)+\frac{3470703125}{36288}\ln(5)-\frac{794596}{945}\gamma-\frac{397298}{945}\ln(u_p)\right. \nonumber\\
&& \left. -\frac{58818333}{1048576}\pi^4+\frac{16293066631}{4718592}\pi^2\right) u_p^7\nonumber\\
&&+\frac{13695499}{47040}\pi u_p^{15/2}\nonumber\\
&&+\left[\frac{497764}{1575}\ln(u_p)^2+\left(\frac{1991056}{1575}\gamma-\frac{66544956203}{3969000}-\frac{11934459}{175}\ln(3)+\frac{195652496}{1575}\ln(2)\right)\ln(u_p)\right. \nonumber\\
&&+\frac{2047686486671407}{13752585000}-\frac{197388844553}{269500}\ln(2)+\frac{359853720161877}{275968000}\ln(3)-\frac{4691575390625}{8515584}\ln(5)\nonumber\\
\nonumber\\
&&-\frac{678223072849}{6082560}\ln(7)-\frac{66544956203}{1984500}\gamma-\frac{23868918}{175}\gamma\ln(3)+\frac{391304992}{1575}\ln(2)\gamma\nonumber\\
&& -\frac{23868918}{175}\ln(2)\ln(3)+\frac{1991056}{1575}\gamma^2+\frac{924796757543}{2013265920}\pi^4\nonumber\\
&&\left. -\frac{11934459}{175}\ln(3)^2-\frac{37216}{15}\zeta(3)+\frac{751271824}{1575}\ln(2)^2+\frac{27703501682741}{19818086400}\pi^2\right] u_p^8\nonumber\\
&&+\frac{1023562537}{1552320}\pi u_p^{17/2}\nonumber\\
&&+\left[-\frac{10161819}{1225}\ln(u_p)^2\right.\nonumber\\
&&+\left(\frac{15523629993}{39200}\ln(3)+\frac{46249898026747}{305613000}-\frac{14486589644}{11025}\ln(2)-\frac{40647276}{1225}\gamma+\frac{3173828125}{14112}\ln(5)\right)\ln(u_p)  \nonumber\\
&&+\frac{46285104644347}{152806500}\gamma-\frac{2067345910491191}{85899345920}\pi^4+\frac{17923252135149887}{1986484500}\ln(2)-\frac{148748447195686881}{25113088000}\ln(3)\nonumber\\
&&-\frac{27786921439609375}{16273281024}\ln(5)+\frac{421370306260043}{219648000}\ln(7)+\frac{1655592}{35}\zeta(3)+\frac{4816187291152031551}{1529593065000}\nonumber\\
&& +\frac{3173828125}{14112}\ln(5)^2-\frac{6286441324}{1225}\ln(2)^2+\frac{3173828125}{7056}\gamma\ln(5)-\frac{40647276}{1225}\gamma^2\nonumber\\
&&+\frac{15523629993}{39200}\ln(3)^2-\frac{1712225112134041}{34681651200}\pi^2+\frac{15523629993}{19600}\gamma\ln(3)-\frac{28973179288}{1025}\ln(2)\gamma\nonumber\\
&& \left. +\frac{3173828125}{7056}\ln(2)\ln(5)+\frac{15523629993}{19600}\ln(2)\ln(3)\right] u_p^9\nonumber\\
&&
+\left(-\frac{137457732402576571}{610248038400}\pi-\frac{40118366}{4725}\pi^3+\frac{4292665162}{165375}\pi\gamma+\frac{2146332581}{165375}\pi\ln(u_p)+\frac{83149713482}{165375}\pi\ln(2)\right. \nonumber\\
&& \left. -\frac{1427220891}{6125}\pi\ln(3)\right) u_p^{19/2}\nonumber
\end{eqnarray}
\begin{eqnarray}
\phantom{\delta U^{e^4}(u_p)}
&& 
+\left[
 -\frac{2438262007}{198450}\ln(u_p)^2+\left(\frac{487959613018}{72765}\ln(2)+\frac{4899895367447}{4584195}+\frac{114443682651}{431200}\ln(3)-\frac{4876524014}{99225}\gamma\right.\right.\nonumber\\
&&\left.
-\frac{40750244140625}{12573792}\ln(5)\right)\ln(u_p)-\frac{20377781024735400904328}{201332687180625}+\frac{9826562433694}{4584195}\gamma\nonumber\\
&&
-\frac{176771772306908307}{276243968000}\ln(3)-\frac{40750244140625}{12573792}\ln(5)^2-\frac{4876524014}{99225}\gamma^2+\frac{114443682651}{431200}\ln(3)^2\nonumber\\
&& 
+\frac{36307823919194}{1403325}\ln(2)^2 +\frac{536631960411}{215600}\ln(2)\ln(3)+\frac{114443682651}{215600}\gamma\ln(3)+\frac{19871612}{63}\zeta(3)\nonumber\\
&&
-\frac{131229423889414613}{8895744000}\ln(7)-\frac{40750244140625}{6286896}\gamma\ln(5)+\frac{1167313947555}{268435456}\pi^6+\frac{975919226036}{72765}\ln(2)\gamma\nonumber\\
&&
+\frac{628897515069490765625}{14499493392384}\ln(5)-\frac{707217483022033957}{1109812838400}\pi^2-\frac{40750244140625}{6286896}\ln(2)\ln(5)\nonumber\\
&&\left. 
+\frac{10992948747002026551}{10995116277760}\pi^4-\frac{476838331512979466}{9833098275}\ln(2)
\right]u_p^{10}\nonumber\\
&& +\left(  \frac{16759623823}{176400}\pi^3+\frac{80294969715785936774437}{45814371482880000}\pi-\frac{6503164207213}{37044000}\pi\ln(u_p)-\frac{100110344015981}{18522000}\pi\ln(2)\right.\nonumber\\
&&\left.
+\frac{2087606910177}{1372000}\pi\ln(3) -\frac{6503164207213}{18522000}\pi\gamma+\frac{206298828125}{296352}\pi\ln(5) \right) u_p^{21/2}
+O_{\ln{}}(u_p^{11})\,.
\end{eqnarray}

The numerical form of this expansion reads
\begin{eqnarray}
\delta U^{e^4}(u_p) &=&-2.0 u_p^2+0.25 u_p^3+83.38296351u_p^4+(737.1849552+48.66666667\ln(u_p))u_p^5\nonumber\\
&& +(2980.049710-120.2 \ln(u_p))u_p^6+429.7389577 u_p^{13/2}\nonumber\\
&& +(19588.97635-420.4211640\ln(u_p)) u_p^7+914.6615445 u_p^{15/2}\nonumber\\
&& +(62630.23815-4853.06274\ln(u_p)+316.0406349\ln(u_p)^2) u_p^8+2071.490767 u_p^{17/2}\nonumber\\
&& +(18432.5611\ln(u_p)-8295.362449\ln(u_p)^2+837868.8305) u_p^9\nonumber\\
&& +(-633183.2616+40773.40995\ln(u_p)) u_p^{19/2}\nonumber\\
&& +(764293.4202\ln(u_p)-12286.53065\ln(u_p)^2+1154095.188) u_p^{10}\nonumber\\
&& +(4816799.276-551514.2235\ln(u_p)) u_p^{21/2} 
+O_{\ln{}}(u_p^{11})
\,.
\end{eqnarray}

Using the relations explicitly written down in  \cite{Tiec:2015cxa} we converted the new information on $\delta U^{e^2}$ and $\delta U^{e^4}$ into a
correspondingly improved knowledge of the EOB potentials $\bar d(x)$ and $q(x)$, namely
\begin{eqnarray}
\label{dPN}
\bar d(x)&=& 6x^2+52x^3+\left(\frac{1184}{15}\gamma-\frac{6496}{15}\ln(2)+\frac{2916}{5}\ln(3)-\frac{23761}{1536}\pi^2-\frac{533}{45}+\frac{592}{15}\ln(x)\right)x^4\nonumber\\
&& +\left(-\frac{2840}{7}\gamma+\frac{120648}{35}\ln(2)-\frac{19683}{7}\ln(3)-\frac{63707}{512}\pi^2+\frac{294464}{175}-\frac{1420}{7}\ln(x)\right)x^5\nonumber\\
&& +\frac{264932}{1575}\pi x^{11/2}\nonumber\\
&& +\left(-\frac{64096}{45}\gamma-\frac{6381680}{189}\ln(2)+\frac{1765881}{140}\ln(3)+\frac{9765625}{2268}\ln(5)+\frac{135909}{262144}\pi^4+\frac{229504763}{98304}\pi^2\right.\nonumber\\
&&\left.
-\frac{31721400523}{2116800}-\frac{32048}{45}\ln(x)\right)x^6\nonumber\\
&& -\frac{21288791}{17640}\pi x^{13/2}\nonumber\\
&&+\left(\frac{4187061434}{99225}\gamma-\frac{876544}{315}\ln(x)\gamma+\frac{8108032}{1575}\ln(2)\ln(x)+\frac{16216064}{1575}\ln(2)\gamma-\frac{3744144}{175}\ln(2)\ln(3)\right.\nonumber\\
&&
-\frac{3744144}{175}\gamma\ln(3)+\frac{18024943666}{496125}\ln(2)+\frac{282753093897}{2156000}\ln(3)+\frac{16384}{3}\zeta(3)-\frac{3091796875}{66528}\ln(5) \nonumber\\
&& 
+\frac{33089536}{1575}\ln(2)^2+\frac{31596265477}{251658240}\pi^4+\frac{3755930660113}{247726080}\pi^2-\frac{876544}{315}\gamma^2-\frac{1872072}{175}\ln(3)^2\nonumber\\
&& \left.
+\frac{629856}{55}\ln(6)-\frac{1340870864165051}{5501034000}-\frac{219136}{315}\ln(x)^2-\frac{1872072}{175}\ln(3)\ln(x)+\frac{2093530717}{99225}\ln(x)\right)x^7\nonumber\\
&&-\frac{1173441809}{3492720}\pi x^{15/2}\nonumber\\
&&+\left(-\frac{281972594008247}{1986484500}\gamma+\frac{232751488}{11025}\ln(x)\gamma-\frac{31370368}{525}\ln(2)\ln(x)-\frac{62740736}{525}\ln(2)\gamma\right.\nonumber\\
&&
+\frac{174802536}{1225}\ln(2)\ln(3)+\frac{174802536}{1225}\gamma\ln(3)+\frac{107340333276983}{283783500}\ln(2)-\frac{25726492389393}{49049000}\ln(3)\nonumber\\
&&
-\frac{1096192}{35}\zeta(3)+\frac{1556814453125}{6054048}\ln(5)+\frac{678223072849}{23166000}\ln(7)-\frac{624682112}{2205}\ln(2)^2+\frac{16273379175661}{1073741824}\pi^4
\nonumber\\
&&
+\frac{2692389474594437}{92484403200}\pi^2 +\frac{232751488}{11025}\gamma^2+\frac{87401268}{1225}\ln(3)^2-\frac{2751525936}{17875}\ln(6)+\frac{58187872}{11025}\ln(x)^2\nonumber\\
&& \left.
+\frac{87401268}{1225}\ln(3)\ln(x)-\frac{831440592970385544103}{440522802720000}-\frac{281464053976247}{3972969000}\ln(x)\right) x^8\nonumber\\
&&
+\left(\frac{144712674728544827}{1678182105600}\pi-\frac{186756088}{33075}\pi\ln(x)+\frac{239421488}{23625}\ln(2)\pi+\frac{3490768}{945}\pi^3-\frac{373512176}{33075}\pi\gamma\right.\nonumber\\
&&\left. 
-\frac{200311704}{6125}\pi\ln(3)\right) x^{17/2}\nonumber\\
&& 
+\left[ -\frac{145060456}{363825}\ln(x)^2+\left(-\frac{4109882910365899}{19423404000}+\frac{2205013489376}{3274425}\ln(2)-\frac{580241824}{363825}\gamma-\frac{215213193}{770}\ln(3)\right.\right.\nonumber\\
&&\left.
-\frac{76708984375}{785862}\ln(5)\right)\ln(x)+\frac{8869707677468340294172589}{188984282366880000}-\frac{4125670253137099}{9711702000}\gamma\nonumber\\
&& 
-\frac{23620001432239865033}{2359943586000}\ln(2)+\frac{18387195312716343}{4932928000}\ln(3)+\frac{1763600530764453125}{1812436674048}\ln(5)-\frac{215213193}{385}\gamma\ln(3)\nonumber\\
&& +\frac{117281890332}{125125}\ln(6)-\frac{7435264}{105}\zeta(3)-\frac{43503165672743}{92664000}\ln(7)+\frac{13438960917574667}{406931374080}\pi^2-\frac{441262176956397691}{1030792151040}\pi^4\nonumber\\
&&
-\frac{215213193}{770}\ln(3)^2-\frac{580241824}{363825}\gamma^2+\frac{5132203667744}{1964655}\ln(2)^2-\frac{76708984375}{785862}\ln(5)^2-\frac{150232915593}{33554432}\pi^6\nonumber\\
&& \left. 
-\frac{76708984375}{392931}\gamma\ln(5)-\frac{215213193}{385}\ln(2)\ln(3)+\frac{4410026978752}{3274425}\ln(2)\gamma-\frac{76708984375}{392931}\ln(2)\ln(5)\right]x^9\nonumber\\
&&+\left(-\frac{10310051408772977303753}{22907185741440000}\pi-\frac{1836704419}{66150}\pi^3+\frac{232145783843}{2315250}\pi\gamma-\frac{83839907743}{771750}\pi\ln(2)\right.\nonumber\\
&&\left.
+\frac{39949476291}{171500}\pi\ln(3)+\frac{232145783843}{4630500}\pi\ln(x)\right) x^{19/2}
+O_{\ln{}}(x^{10})
\end{eqnarray}
and
\begin{eqnarray}
\label{qPN}
q(x)&=& 8x^2+\left(\frac{496256}{45}\ln(2)-\frac{33048}{5}\ln(3)-\frac{5308}{15}\right) x^3\nonumber\\
&&
+\left(\frac{10856}{105}\gamma-\frac{40979464}{315}\ln(2)+\frac{14203593}{280}\ln(3)+\frac{9765625}{504}\ln(5)-\frac{93031}{1536}\pi^2+\frac{1295219}{350}+ \frac{5428}{105}\ln(x)\right) x^4\nonumber\\
&&+ \frac{88703}{1890}\pi x^{9/2}\nonumber\\
&& +\left(-\frac{617716}{315}\gamma-\frac{308858}{315}\ln(x)+\frac{65887036}{63}\ln(2)-\frac{36073593}{112}\ln(3)-\frac{8787109375}{27216}\ln(5)+\frac{81030481}{65536}\pi^2\right.\nonumber\\
&&\left.
+\frac{559872}{7}\ln(6)-\frac{7518451741}{1270080}\right) x^5\nonumber\\
&&-\frac{714117331}{846720}\pi x^{11/2}\nonumber\\
&& +\left(\frac{138169844888}{1819125}\gamma+\frac{69084922444}{1819125}\ln(x)-\frac{3250526464}{4725}\ln(2)\gamma+\frac{13728528}{35}\ln(2)\ln(3)+\frac{13728528}{35}\gamma\ln(3)\right.\nonumber\\
&&
-\frac{527856862616}{16372125}\ln(2)-\frac{12960490645107}{6899200}\ln(3)+\frac{25344}{5}\zeta(3)+\frac{27397616796875}{9580032}\ln(5)+\frac{678223072849}{2280960}\ln(7)\nonumber\\
&& 
-\frac{2065918336}{1575}\ln(2)^2-\frac{109837713789}{83886080}\pi^4+\frac{1463044337673}{91750400}\pi^2-\frac{451968}{175}\gamma^2+\frac{6864264}{35}\ln(3)^2-\frac{579887424}{385}\ln(6)\nonumber\\
&&\left. 
-\frac{451968}{175}\ln(x)\gamma+\frac{6864264}{35}\ln(3)\ln(x)-\frac{939101654498857}{3056130000}-\frac{112992}{175}\ln(x)^2-\frac{1625263232}{4725}\ln(2)\ln(x)\right) x^6\nonumber\\
&& +\frac{226615901761}{167650560}\pi x^{13/2}\nonumber\\
&& +\left(-\frac{29186389360543}{36786750}\gamma-\frac{3173828125}{2646}\ln(2)\ln(5)-\frac{3173828125}{2646}\gamma\ln(5)-\frac{9440966259}{4900}\ln(3)\ln(x)\right.\nonumber\\
&&
+\frac{322866894016}{33075}\ln(2)\gamma-\frac{9440966259}{2450}\ln(2)\ln(3)-\frac{9440966259}{2450}\gamma\ln(3)-\frac{97783791533166503}{2979726750}\ln(2) \nonumber\\
&& 
+\frac{87139874452615209}{6278272000}\ln(3)-\frac{2452928}{35}\zeta(3)-\frac{2899973891640625}{452035584}\ln(5)-\frac{9257841833399257}{1482624000}\ln(7)\nonumber\\
&& +\frac{210393017888}{11025}\ln(2)^2+\frac{19047555410493}{10737418240}\pi^4-\frac{3975430726567129}{92484403200}\pi^2+\frac{168910688}{3675}\gamma^2-\frac{9440966259}{4900}\ln(3)^2\nonumber\\
&&
+\frac{2573147182608}{175175}\ln(6)-\frac{3173828125}{5292}\ln(5)^2-\frac{3173828125}{5292}\ln(5)\ln(x)+\frac{161433447008}{33075}\ln(2)\ln(x)\nonumber\\
&& \left.
+\frac{690294961714478265797}{293681868480000}-\frac{29186389360543}{73573500}\ln(x)+\frac{42227672}{3675}\ln(x)^2+\frac{168910688}{3675}\ln(x)\gamma  \right)x^7\nonumber\\
&& +\left(\frac{8192870254937920639}{30981823488000}\pi-\frac{15200768606}{11025}\ln(2)\pi-\frac{11631519958}{496125}\pi\gamma+\frac{4598822871}{6125}\pi\ln(3)+\frac{108705794}{14175}\pi^3\right.\nonumber\\
&&\left.
-\frac{5815759979}{496125}\pi\ln(x)\right) x^{15/2}\nonumber\\
&& 
+\left[\frac{131228022231920707}{196661965500}\gamma+\frac{23800697662770506993}{65553988500}\ln(2)-\frac{4958146688407013943}{39463424000}\ln(3)\right.\nonumber\\
&&
-\frac{4492372832662738703125}{43498480177152}\ln(5)-\frac{2906254027437804}{67442375}\ln(6)+\frac{1519264}{315}\zeta(3)+\frac{710656240002840019}{10674892800}\ln(7)\nonumber\\
&& 
+\frac{472332484052074531}{678218956800}\pi^2+\frac{9617337404302759049}{24739011624960}\pi^4+\frac{105898193359375}{9430344}\ln(5)^2+\frac{226550022075}{33554432}\pi^6\nonumber\\
&& -\frac{188687137328}{1091475}\gamma^2-\frac{7036471296}{2695}\ln(6)^2+\frac{181114018983}{15400}\ln(3)^2-\frac{4527732156900112}{29469825}\ln(2)^2\nonumber\\
&&
-\frac{85388056818784}{1091475}\ln(2)\gamma-\frac{14072942592}{2695}\ln(2)\ln(6)+\frac{105898193359375}{4715172}\gamma\ln(5)-\frac{14072942592}{2695}\gamma\ln(6)\nonumber\\
&& 
+\frac{181114018983}{7700}\gamma\ln(3)+\frac{105898193359375}{4715172}\ln(2)\ln(5)+\frac{181114018983}{7700}\ln(2)\ln(3)
-\frac{7190610346934768219939609}{161986527743040000}\nonumber\\
&& +\left(-\frac{7036471296}{2695}\ln(6)+\frac{105898193359375}{9430344}\ln(5)+\frac{131815385968880707}{393323931000}-\frac{188687137328}{1091475}\gamma+\frac{181114018983}{15400}\ln(3)\right.\nonumber\\
&&\left.\left.
-\frac{42694028409392}{1091475}\ln(2)\right)\ln(x)-\frac{47171784332}{1091475}\ln(x)^2\right] x^8\nonumber
\end{eqnarray}
\begin{eqnarray}
\phantom{q(x)}
\label{qPN}
&&+\left(-\frac{309249455540719514934031}{84580378122240000}\pi-\frac{206298828125}{111132}\pi\ln(5)-\frac{431496991403}{3175200}\pi^3+\frac{52009951116491}{111132000}\pi\gamma\right. \nonumber\\
&& \left.
+\frac{88244053021571}{4445280}\pi\ln(2)-\frac{21764539709991}{2744000}\pi\ln(3)+\frac{52009951116491}{222264000}\pi\ln(x)\right) x^{17/2}
+O_{\ln{}}(x^{9})\,.
\end{eqnarray}

As mentioned above, another useful dynamical function is the  precession function $\rho(u)$ introduced in \cite{Damour:2009sm} and related there
to the EOB potentials $a(u)$ and $\bar d(u)$. Namely (denoting the argument of the function $\rho$ as $x$)
\beq
\label{rhoeob1}
\rho(x)=\rho_E(x)+\rho_a(x)+\rho_{\bar d}(x)\,,
\eeq
with
\begin{eqnarray}
\label{rhoeob2}
\rho_E(x)&=& 4x \left(1-\frac{1-2x}{\sqrt{1-3x}}\right)\,,\nonumber\\
\rho_a(x)&=& a(x)+xa'(x)+\frac12 x (1-2x)a''(x)\,,\nonumber\\
\rho_{\bar d}(x)&=&(1-6x) \bar d(x) \,.
\end{eqnarray}
We then find

\begin{eqnarray}
\label{rhoPN}
\rho(x)&=&14 x^2+\left(\frac{397}{2}-\frac{123}{16}\pi^2\right) x^3+\left(\frac{5024}{15}\gamma-\frac{215729}{180}+\frac{2512}{15}\ln(x)+\frac{2916}{5}\ln(3)+\frac{1184}{15}\ln(2)+\frac{58265}{1536}\pi^2\right) x^4\nonumber\\
&&+\left(\frac{27824}{35}\ln(2)-\frac{6325051}{800}+\frac{1135765}{1024}\pi^2-\frac{202662}{35}\ln(3)-\frac{22672}{7}\gamma-\frac{11336}{7}\ln(x)\right) x^5\nonumber\\
&&+\frac{199876}{315}\pi x^{11/2}\nonumber\\
&&+\left(\frac{4990303259}{589824}\pi^2-\frac{256727518799}{6350400}+\frac{435213}{20}\ln(3)+\frac{3606884}{945}\gamma-\frac{37648124}{945}\ln(2)+\frac{1803442}{945}\ln(x)\right.\nonumber\\
&&\left.
-\frac{7335303}{32768}\pi^4+\frac{9765625}{2268}\ln(5)\right) x^6\nonumber\\
&&-\frac{1429274}{225}\pi x^{13/2}\nonumber\\
&&+\left(-\frac{3725312}{1575}\ln(2)^2-\frac{419921875}{6048}\ln(5)-\frac{3744144}{175}\gamma\ln(3)-\frac{3744144}{175}\ln(2)\ln(3)+\frac{253952}{15}\zeta(3)\right.\nonumber\\
&&
-\frac{13586432}{1575}\ln(x)\gamma-\frac{230019793907682883}{440082720000}-\frac{1872072}{175}\ln(3)\ln(x)-\frac{20598784}{1575}\ln(2)\gamma+\frac{12659060941523}{1238630400}\pi^2\nonumber\\
&&
+\frac{681396625634}{5457375}\gamma+\frac{229716339147}{2156000}\ln(3)-\frac{3396608}{1575}\ln(x)^2+\frac{1823766172754}{5457375}\ln(2)-\frac{10299392}{1575}\ln(2)\ln(x)\nonumber\\
&& \left. +\frac{471044952937}{251658240}\pi^4+\frac{340698312817}{5457375}\ln(x)-\frac{1872072}{175}\ln(3)^2-\frac{13586432}{1575}\gamma^2\right) x^7\nonumber\\
&&+\frac{18719967989}{1455300}\pi x^{15/2}\nonumber\\
&&+\left(-\frac{82814168955181}{132432300}\ln(2)+\frac{36686848}{441}\ln(x)\gamma+\frac{1920044921875}{4036032}\ln(5)+\frac{2269129471514627499419}{176209121088000}\right.\nonumber\\
&&
+\frac{148969692}{1225}\ln(3)^2+\frac{36686848}{441}\gamma^2+\frac{103653376}{3675}\ln(2)\gamma+\frac{9171712}{441}\ln(x)^2 +\frac{297939384}{1225}\gamma\ln(3)\nonumber\\
&&
-\frac{948480}{7}\zeta(3)+\frac{297939384}{1225}\ln(2)\ln(3)+\frac{678223072849}{23166000}\ln(7)+\frac{148969692}{1225}\ln(3)\ln(x)\nonumber\\
&&+\frac{51826688}{3675}\ln(2)\ln(x)-\frac{6517218707007553}{55490641920}\pi^2-\frac{1442495323220011}{3972969000}\ln(x)\nonumber\\
&&\left.
-\frac{557542163367261}{392392000}\ln(3)-\frac{1444834607367211}{1986484500}\gamma-\frac{2049476608}{11025}\ln(2)^2-\frac{626168320805261}{5368709120}\pi^4\right) x^8\nonumber
\end{eqnarray}
\begin{eqnarray}
\phantom{\rho(x)}
\label{rhoPN}
&&
+\left(\frac{105699126344597143}{524431908000}\pi-\frac{4707645616}{165375}\pi\gamma-\frac{200311704}{6125}\pi\ln(3)-\frac{4004219056}{165375}\ln(2)\pi-\frac{2353822808}{165375}\pi\ln(x)\right.\nonumber\\
&&\left.
+\frac{43996688}{4725}\pi^3\right) x^{17/2}\nonumber\\
&&+\left[-\frac{42935456848}{1091475}\ln(x)^2+\left(-\frac{1995219783}{3850}\ln(3)-\frac{20475902395612297}{262215954000}-\frac{171741827392}{1091475}\gamma+\frac{2444046775616}{3274425}\ln(2)\right.\right.\nonumber\\
&&\left.
-\frac{76708984375}{785862}\ln(5)\right)\ln(x)-\frac{20657146063017097}{131107977000}\gamma+\frac{10969454340865467}{2466464000}\ln(3)+\frac{226626361596}{125125}\ln(6)\nonumber\\
&&
-\frac{76708984375}{785862}\ln(5)^2-\frac{171741827392}{1091475}\gamma^2-\frac{1995219783}{3850}\ln(3)^2+\frac{6754948737728}{1964655}\ln(2)^2\nonumber\\
&&-\frac{1995219783}{1925}\gamma\ln(3)-\frac{76708984375}{392931}\gamma\ln(5)+\frac{3956569170916362731724487183}{18142491107220480000}+\frac{8519104}{315}\zeta(3)\nonumber\\
&&
-\frac{29163592132507}{46332000}\ln(7)-\frac{76708984375}{392931}\ln(2)\ln(5)+\frac{4888093551232}{3274425}\ln(2)\gamma-\frac{64674832921484375}{906218337024}\ln(5)\nonumber\\
&& 
+\frac{3466357618648439}{27128758272}\pi^2+\frac{128148402261}{16777216}\pi^6-\frac{1995219783}{1925}\ln(2)\ln(3)-\frac{6269062781928031361}{2748779069440}\pi^4\nonumber\\
&&\left.
-\frac{2702219779688690213}{235994358600}\ln(2)  \right]x^9\nonumber\\
&& +\left(\frac{174754006268}{1157625}\pi\ln(x)-\frac{2849519528}{33075}\pi^3+\frac{349508012536}{1157625}\pi\gamma\right.\nonumber\\
&&\left.
+\frac{16864298172}{42875}\pi\ln(3)-\frac{2195209992943672765961}{1431699108840000}\pi+\frac{51802382504}{385875}\pi\ln(2)\right) x^{19/2}
+O_{\ln{}}(x^{10})\,.
\end{eqnarray}

\end{widetext}

\section{Estimating the order of magnitude of the coefficients of PN expansions}

Before comparing the numerical values of these 9.5PN-accurate functions to corresponding published numerical SF estimates \cite{Barack:2010ny,Akcay:2015pjz}, it is useful to have at hand a rough estimate of the theoretical error 
associated with such PN-expanded functions.  
We shall do this via two complementary approaches. Our first estimate will follow the spirit of Section IV in Ref.  \cite{Bini:2014zxa}. 
The idea there was to use the existence of a power-law singularity
at the lightring \cite{Akcay:2012ea} of the various SF or EOB potentials to estimate, for a given potential  $f(u) = \sum_{n<N} f_n u^n + \epsilon_{f}^{N}(u)$,
both the order of magnitude of the PN expansion coefficients $f_n$, and that of the Nth PN remainder $\epsilon_{f}^{N}(u) =O_f(u^N)$, from the knowledge of its  lightring singularity. 
The coarsest such estimate consists in saying that the radius of convergence of
a power series\footnote{Here, we formally proceed as if the PN expansion contained only integer powers. The existence of
logarithmic corrections, starting at 4PN \cite{Damour:2009sm,Blanchet:2010zd}, and of a sub-series, starting at 5.5PN \cite{Shah:2013uya,Bini:2013rfa,Blanchet:2013txa}, containing half-integer powers, indicates that, from a
theoretical point of view, a more subtle treatment should be applied. See below for the logarithmic corrections.}, 
$\sum_N f_N u^N$, is determined by the location of the singularity closest to the origin in the complex $u$ plane. Assuming that the closest singularity 
is the lightring one at $u=1/3$ determines the radius of convergence as being $|u|_{\rm conv}= \frac13$. This simple
consideration tells us that the large--$N$ asymptotic values of the Taylor expansion coefficients $f_N$ is of order
\beq
f_N \sim 3^N\,.
\eeq
One can, however, refine this exponential estimate by power-law corrections in $N$. 
Indeed, given a certain function $f(u) = \sum_N f_N u^N$, its first derivative with respect to (wrt) $u$
will be $f'(u)= \sum_N  N f_{N} u^{N -1}$, so that $(f')_N = (N+1) f_{N+1}$. In other words, each derivative
adds an asymptotic factor $N$ to the growth of the $f_N$'s.
For instance, the existence in EOB theory of the link \eqref{rhoeob1}, \eqref{rhoeob2} between the precession function
$\rho(u)$ and the first two derivatives of the primary EOB radial (1SF) potential $a(u)$ suggests that, asymptotically,
\beq
\label{rhoN1}
\rho_N \sim  N^2  a_N\,,
\eeq
where $a_N$ are the PN expansion coefficients of $a(u)$ and $\rho_N$ those of $\rho(u)$. [Here, we assume that
the PN coefficients $\bar d_N$ of $\bar d(u)$  do not cancel the growth with $N$ entailed by the two derivatives
in the first equation  \eqref{rhoeob2}. Our numerical studies below will confirm this assumption.]

There is an alternative perspective on the additional power-law growth (of the type of the factor $N^2$ in \eqref{rhoN1}).
It consists in using more information about the singularity structure of the considered function $f(u)$ near its closest
singularity. Indeed, if we knew, for instance, that $f(u)$ had a power-law singularity near $u=\frac13$ of the type
\beq
\label{sing}
f^{\rm sing}(u) = K_f (1- 3u)^{- n_f}\,,
\eeq
($K_f$ denoting a constant), we would expect\footnote{This expectation is based on the usual integral Cauchy formula giving the coefficients of the Laurent expansion of an analytic function. By deforming the contour of integration so that it gets near the (closest) singularity one sees
that the Cauchy integral can be approximated by an analogous integral involving  $ f^{\rm sing}(u) $.} the expansion coefficients $f_N$ of $f(u)$ to be asymptotically approximated by the
expansion coefficients of its singular piece \eqref{sing}, namely
\beq
\label{expsing}
f^{\rm sing}_N =K_f  \, \binom{- n_f}{N} \, 3^N \approx C_f \, N^{n_f - 1} \, 3^N\,,
\eeq
with $C_f= K_f/ \Gamma(n_f)$. Here we see that while the location of the singularity determines the
exponentially growing factor $3^N$, the sub-leading power-law growth $\propto N^{n_f - 1}$ would be
determined by the power $-n_f$ of the singular piece \eqref{sing}.
Consistently with our remarks above, note that acting on $f(u)$ by $k$ derivatives changes $n_f$ into $n_f+k$,
and correspondingly increases the power-law growth of the $f_N$'s by $+ k$.

Ref.  \cite{Akcay:2012ea} has found that the lightring singularity structure of the basic 1SF EOB potential $a(u)$ was
$ a^{\rm sing}(u) = K_a (1- 3u)^{- n_a}$  with $K_a \simeq \frac14$ and $n_a=\frac12$.  
One would then expect a large-$N$ behavior $a_N \simeq C_a  N^{- 1/2}  3^N$ 
with $C_a \simeq 1/(4 \Gamma(\frac12)) \simeq  0.14 $. We studied the evolution with $N$ of the
PN coefficients $a_N$ of $a(u)$ by using the available high-PN results of Refs. \cite{Bini:2013rfa,Kavanagh:2015lva}.
We confirmed the basic exponential growth $a_N \sim 3^N$. Indeed, Table \ref{hata} in Appendix \ref{appTables}
 displays, in its first column, the values of the rescaled PN coefficients $\widehat a_N \equiv a_N/3^N$ 
 [with all logarithms replaced by  $\ln(\frac13)$; see below for the logarithmic dependence]. 
 These rescaled coefficients  are seen to
remain (roughly) of order unity (in absolute magnitude), even up to $N= 23$ for which $3^{23}= 0.941432 \times 10^{11}$.
More precisely, we have 
$ 0.1 \lesssim \widehat a_N \lesssim 1$, when $N$ varies between $3$ and $20$, while, for $ N=21, 22, 23$, we have
$|\widehat a_N| \approx 2.254, 1.459, 3.313$, respectively. [We do not know if the fact that the latter values are slightly larger than 1 signals the beginning of a growth for very large $N$'s.] We did not see any sign of the expected mild decay  $\widehat a_N \simeq C_a  N^{- 1/2}$.
This might be due to the more complicated singularity structure (beyond the leading-order  power-law) found in \cite{Akcay:2012ea},
or to the fact that the $N^{- 1/2}$ behavior sets in only for very large $N$'s. [Note also that, after having factored
 the clear $3^N$ growth, the rescaled coefficients $\widehat a_N$ behave rather erratically, and do not show
 any sign of converging towards a simple behavior.]
 If we were only interested in estimating the PN error for values of $ u$ in the strong-field domain,
 and for values of $N$ around 10, we could simply use the simple estimate 
 $a_N \simeq C_a 3^N$ with $|C_a| \sim 1$.
However, as we are also interested in knowing what happens when  $u \ll 0$, we should remember that
the PN expansion coefficients run logarithmically with $u$ as $u \to 0$. We therefore kept the full available 
high PN information \cite{Bini:2013rfa,Kavanagh:2015lva} to also investigate the effect of the logarithms.
It is technically convenient to work with the rescaled independent variable
$u_3 \equiv 3 u$ (with respect to which the singularity is located at $u_3=1$), and expand $a(u) \equiv a_3(u_3)$ in powers of $u_3$ 
\beq
\label{au3pnexp}
a(u) = \sum_N (\widehat a_N  + \widehat a'_N  \ln(u_3) + \widehat a''_N  (\ln(u_3))^2 + \cdots ) \, u_3^N\,.
\eeq
Here, the $\widehat a_N$ are the same (rescaled) coefficients as above (obtained by replacing $\ln u$ by $\ln \frac13$).
The higher logarithmic coefficients  $\widehat a'_N ,  \widehat a''_N , \cdots$ are displayed in the other columns of
Table \ref{hata} in Appendix \ref{appTables}.
We see that the first  logarithmic (rescaled) coefficients  $ \widehat a'_N$ are either comparable to the $ \widehat a_N$,
or slightly smaller in absolute value (the signs, as well as the relative signs, of $\widehat a_N$ and $ \widehat a'_N $
fluctuate). The higher logarithmic terms   appear only at  $u^8$, and their coefficients $\widehat a''_N , \cdots$
are found to be generally smaller. We shall neglect them in the following.
 Going back to the original PN-expansion coefficients $a_N(u)$ of $a(u) = \sum_N a_N(u) u^N$, 
 we can then write their combined $N$ and $u$ dependence as (neglecting higher logarithms)
\beq
\label{aN}
a_N(u) \simeq ( \widehat a_N + \widehat a'_N  \ln(3 u) ) \, 3^N\,; 
\eeq
with  $|\widehat a_N| \sim |\widehat a'_N | \sim 1$.
Then, in view of \eqref{rhoeob2}, we expect a corresponding approximate asymptotic behavior of the
PN expansion coefficients of the precession function $\rho(u)$ of the type
\beq
\label{rhoN1new}
\rho_N(u) \simeq  ( \widehat \rho_N + \widehat \rho'_N  \ln(3 u) ) \, 3^N  ; 
\eeq
with $ |\widehat \rho_N | \sim |\widehat \rho'_N | \sim N^2$.

We tested this expectation on the  9.5PN-accurate expansion of $\rho(u)$ given above.
An $N^2$ scaling seems to be in reasonable agreement with the currently known PN coefficients,
and we found (using a $(N-1)^2$ scaling and relying on the 8PN and 9PN nonlogarithmic coefficients to fix the overall coefficient) for the coefficients of $\rho$
(see Table \ref{hatrho} in Appendix \ref{appTables})
\beq
\label{rhoN2}
 |\widehat \rho'_N | \lesssim |\widehat \rho_N | \sim 2.5 \, (N -1)^2 \, .
\eeq
A similar study of the PN expansion coefficients of the function $\bar d(u)$ (see Table \ref{hatbard} in Appendix \ref{appTables})
leads to a growth similar to the case of the function $\rho(u)$, with simply a slightly smaller
overall coefficient, i.e.
\beq
\label{dN1}
\bar d_N (u) \simeq  ( \widehat {\bar d}_N + \widehat {\bar d}'_N  \ln(3 u) ) \, 3^N  ; 
\eeq
with
\beq
\label{dN2}
 |\widehat {\bar d}'_N | \lesssim |\widehat {\bar d}_N | \sim 0.7 \, (N -1)^2 \, .
\eeq
Finally,  the fact that the EOB potential $q(u)$ is related to the redshift coefficient $\delta U^{e^0}(u) \sim a(u)$  by four derivatives \cite{Tiec:2015cxa} suggests, in view of the argument above, that
\beq
\label{}
q_N \sim  N^4  a_N \, .
\eeq
We tested this expectation on the  9.5PN-accurate expansion of $q(u)$ given above.
An $N^4$ [or $(N-1)^4$] scaling seems to be in reasonable agreement with the currently known PN coefficients,
and we found (using mainly the last point for numerically estimating the coefficient; see Table \ref{hatq} in Appendix \ref{appTables})
\beq
\label{qN1}
q_N(u) \simeq  ( \widehat  q_N + \widehat q'_N  \ln(3 u) ) \, 3^N\,; 
\eeq
with
\beq
\label{qN2}
 |\widehat q'_N | \lesssim |\widehat q_N | \sim 0.26 \, (N -1)^4 \,.
\eeq

Let us now turn to estimating the Nth PN remainder $\epsilon_{f}^{N}(u) =O_f(u^N)$
in the PN expansion of some potential: $f(u) = \sum_{n<N} f_n u^n + \epsilon_{f}^{N}(u)$.
 As mentioned in \cite{Bini:2014zxa}, the remainder will share the singularity structure of $f(u)$ 
 and will therefore ultimately blow up as $ f^{\rm sing}(u) = K_f (1- 3u)^{- n_f}$ near the lightring.
 However, if one is interested (as we are here) in estimating the PN remainder at values of $u$ significantly away
 from the singular point, we can neglect any additional factor $\propto (1- 3u)^{- n_f}$ and simply estimate the
 remainder by the expected (unknown) next term in the PN expansion. Using our results above on the growth with $N$
 of the PN expansion coefficients, we can estimate (by extrapolation on the value of $N$) that the theoretical
 errors on our 9.5PN expansions Eqs. \eqref{dPN}, \eqref{qPN} and \eqref{rhoPN}  are, respectively,
 \beq
 \Delta \rho^{9.5\rm PN}\sim \rho_{10}(u) \,,
 \eeq
 with
 \begin{eqnarray}
\label{err_rho}
  |\rho_{10}(u)| &\lesssim& (1+ |\ln( 3u)| \, ) [2.5 \, (N -1)^2 (3 u)^N]_{N=10} \nonumber \\
  &\simeq& 203 (1+  | \ln( 3u)| \, ) (3 u)^{10}\,.
 \end{eqnarray}
 Though, as discussed above, and as
can be seen in Tables \ref{hata}--\ref{hatq} in Appendix \ref{appTables}, there is some indication that the coefficient of $\ln(3u)$
is often smaller than that of the nonlogarithmic term, we have conservatively 
preferred, in estimating an upper bound on the theoretical error,
to assume a relative coefficient equal to one for the logarithmic term. 
If one were to relax this conservative assumption,  one could replace in
Eq. \eqref{err_rho} [as well as in Eqs. \eqref{err_bard}, \eqref{err_q} below] the factor $1 + |\ln(3u)|$  by a factor $1 +
c \, |\ln(3u)|$ with some positive coefficient  $c<1$ (e.g. $c\sim \frac12$). [Note that, as we are only interested in the domain
$ u < \frac13$, we could replace $| \ln( 3u)|$ by $- \ln( 3u)$.]

We have tested the reasonableness of the estimate \eqref{err_rho} by inserting in the $a(u)$-related contribution, $\rho_a(u)$,
to $\rho(u)$ [see Eqs.  \eqref{rhoeob1}, \eqref{rhoeob2}] the
known 10PN-accurate value of $a(u)$ straightforwardly computed from the results of Ref. \cite{Kavanagh:2015lva} for $\delta U^{e^0}$. We found (numerically)
\begin{eqnarray}
\rho_a^{\rm 10PN} &\approx& \left[112.44232 + 18. 94528 \ln(3u) - 4.08314 (\ln(3u) )^2\right. \nonumber\\
&& \left. + 0.13206 (\ln(3u) )^3\right] (3u)^{10},
\end{eqnarray}
which confirms the order of magnitude of our estimate of $\rho_{10}(u)$. 

Similarly, we get as estimate of the PN error on our 9.5PN expansion of $\bar d(u)$
\beq
 \Delta {\bar d}^{9.5\rm PN}\sim {\bar d}_{10}(u) \, ,
 \eeq
 with
 \beq
\label{err_bard}
 |{\bar d}_{10}(u)| \lesssim 57  (1+ |\ln( 3u)| \, ) (3 u)^{10} \, .
 \eeq
Finally, the corresponding estimate for the 9.5PN expansion of $ q(u)$ (which goes up to $N=8.5$) reads
\beq
 \Delta { q}^{9.5\rm PN}\sim {q}_{9}(u) \, ,
 \eeq
 with
 \beq
\label{err_q}
 |{q}_{9}(u)| \lesssim 1065  (1+ |\ln( 3u)| \, ) (3 u)^{9} \, .
 \eeq

\section{Comparing 9.5PN-accurate theoretical results to self-force numerical data}

In the present section we shall compare our 9.5PN-accurate theoretical results to corresponding SF numerical data.
First, we display in Table \ref{tab:2} below the numerical values of the 9.5PN-accurate expansions of the functions
$\rho(u_p), \bar d (u_p)$ and $q(u_p)$  for selected values of the  semi-latus rectum $p \equiv 1/u_p$. These numerical
values are given with twelve significant digits. In addition, for the last four entries, we have also given (on a second line)
the digits that our estimated PN error suggests as being meaningful (the PN error being indicated as a last digit,
within parentheses).

We  then use the theoretical estimates, given in the preceding section, of the PN errors on 
$\rho^{9.5 \rm PN}$, $\bar d^{9.5 \rm PN}$ and  $ q^{9.5 \rm PN}$ to gauge the agreement between
these 9.5PN-accurate expansions and some of the currently published
corresponding numerical SF estimates, namely:  \cite{Barack:2010ny} for \footnote{While writing up this paper we were informed by Maarten van de Meent
that he is finalizing much more accurate numerical computations of $\rho(u)$ \cite{vandemeent16}.}  $\rho(u)$,  and \cite{Akcay:2015pjz} for $\bar d(u)$, and $q(u)$.

Our comparisons are displayed in Tables \ref{tab:3}, \ref{tab:4} and \ref{tab:5}. Each Table displays successively: $p \equiv 1/u_p$, the difference $f^{\rm num}(u_p) - f^{9.5 \rm PN}(u_p)$ between the
numerical estimate $f^{\rm num}(u_p)$ and our analytical one  $f^{9.5 \rm PN}(u_p)$, the numerical error estimate $ \Delta f^{\rm num}$,
the analytical one $ \Delta f^{9.5 \rm PN }$,
and finally the ratio  $[ f^{\rm num}(u_p) - f^{9.5 \rm PN}(u_p) ]/ {\rm sup}(\Delta f^{9.5 \rm PN},  \Delta f^{\rm num})$ . 
The latter ratios are decorated with a star when the maximum (estimated)
error is of numerical origin.   

The fact that the un-starred ratios in the last column are, with very few exceptions\footnote{The only exceptions are:
$p=75$ for $q$ and $p=25$ for $\rho$. For these values the numerical and analytical errors are comparable. Maybe one of the
errors is underestimated.},
 smaller than one
 confirms the correctness of our analytical results. [The fact that they are often
 $\lesssim 0.1$ suggests we overestimated the theoretical error.] Note also 
  that the starred ratios (those for which it is the numerical error which dominates)
 are, with very few exceptions\footnote{The only exceptions are:
$p=100$ for $q$ and $p=30$ for $\rho$. Maybe one of the
errors has been underestimated.}, smaller than one.

\begin{widetext}

\begin{table}
\centering
\caption{Numerical values for $\rho$, $\bar d$ and $q$ from our 9.5PN expressions.}
\begin{ruledtabular}
\begin{tabular}{llll}
$p=1/u_p$ & $\rho^{9.5 \rm PN}$ &${\bar d}^{9.5 \rm PN} $ & $q^{9.5 \rm PN}$\cr
\hline
6.5& 0.671764919305 & 0.486993641405& 0.182950449073\cr
7& 0.561736650217 & 0.382787591949& 0.198169550839\cr
8& 0.406156070112 & 0.252398213123& 0.169101047037\cr 
9& 0.305647401966 & 0.177904676295& 0.133125987258\cr
10& 0.237602128640 & 0.131733356706&     0.105217407767\cr
12& 0.154573521964 &0.0801707241916& 0.0696067840355\cr
14& 0.108052490472    &0.0536892308992& 0.0493762760810\cr
16& 0.0795303790344& 0.0383704313535& 0.0368759609969\cr 
18& 0.0608514155206& 0.0287425071020& 0.0286070525638\cr
20& 0.0479869269632& 0.0223099143151& 0.0228464642737\cr
50& 0.649373236809$\times 10^{-2}$& 0.282840892030$\times 10^{-2}$& 0.339474535246$\times 10^{-2}$\cr 
& 0.0064937324(5) & 0.0028284089(1) & 0.00339475(4) \cr
75& 0.275991903932$\times 10^{-2}$& 0.119177507244$\times 10^{-2}$& 0.148216610087$\times 10^{-2}$\cr 
& 0.002759919039(9) & 0.001191775072(3) & 0.001482166(1) \cr
100&  0.151587749099$\times 10^{-2}$& 0.652456222249$\times 10^{-3}$& 0.825891422094$\times 10^{-3}$\cr 
& 0.0015158774910(5) & 0.0006524562222(2) & 0.00082589142(9)\cr
1000 &0.141215418609$\times 10^{-4}$& 0.605194989392$\times 10^{-5}$& 0.802826796636$\times 10^{-5}$\cr
& 0.00001412154186085476269(8) & 0.00000605194989391653304(2) & 0.0000080282679663565(1)\cr
\end{tabular}
\end{ruledtabular}
\label{tab:2}
\end{table}

\begin{table}
\centering
\caption{Difference between our 9.5PN results on $\bar d(u_p)$ and a sample of numerical values from Ref. \cite{Akcay:2015pjz}.}
\begin{ruledtabular}
\begin{tabular}{lllll}
$p=1/u_p$ & $\bar d^{\rm num} -\bar d^{9.5\rm PN}$ & $\Delta \bar d^{\rm num}$ & $\Delta {\bar d}^{9.5\rm PN}$&  $(\bar d^{\rm num} -\bar d^{9.5\rm PN})/{\rm sup}(\Delta {\bar d})$ \cr
\hline
\hline                                                                    
6.25&  0.1958$\times 10^{-1}$& 0.38$\times 10^{-5}$&     0.64$\times 10^{-1}$& 0.31\cr 
7.5&   0.2837$\times 10^{-2}$& 4.5$\times 10^{-8}$&      0.11$\times 10^{-1}$& 0.25\cr 
9.375& 0.2638$\times 10^{-3}$& 1.4$\times 10^{-9}$&      0.14$\times 10^{-2}$&    0.19\cr 
12&    0.1814$\times 10^{-4}$& 2.0$\times 10^{-10}$& 0.13$\times 10^{-3}$&   0.14\cr
15&    0.1477$\times 10^{-5}$& 1.2$\times 10^{-10}$&     0.15$\times 10^{-4}$&      0.097\cr 
20&    0.4312$\times 10^{-7}$& 7.1$\times 10^{-11}$&  9.5$\times 10^{-7}$& 0.045\cr 
25&    0.7734$\times 10^{-9}$& 4.8$\times 10^{-11}$& 1.1$\times 10^{-7}$& 0.0070\cr 
30&   -0.4376$\times 10^{-9}$& 3.5$\times 10^{-11}$& 1.9$\times 10^{-8}$& -0.023\cr 
40&   -0.8033$\times 10^{-10}$& 2.1$\times 10^{-11}$&  1.2$\times 10^{-9}$&   -0.070\cr 
50&   -0.1030$\times 10^{-10}$& 1.5$\times 10^{-11}$& 1.3$\times 10^{-10}$& -0.078\cr 
60&   -0.5004$\times 10^{-11}$& 1.1$\times 10^{-11}$& 2.2$\times 10^{-11}$&  -0.22\cr
75&   -0.4399$\times 10^{-12}$& 7.2$\times 10^{-12}$& 2.5$\times 10^{-12}$& -0.061${}^*$\cr 
100&  -0.2494$\times 10^{-12}$& 4.4$\times 10^{-12}$&  1.5$\times 10^{-13}$& -0.057${}^*$\cr 
200&  -0.3060$\times 10^{-12}$& 1.3$\times 10^{-12}$& 1.7$\times 10^{-16}$&  -0.24${}^*$\cr 
400&   0.2815$\times 10^{-13}$& 4.0$\times 10^{-13}$& 1.9$\times 10^{-19}$&   0.070${}^*$\cr
\end{tabular}
\end{ruledtabular}
\label{tab:3}
\end{table}

\begin{table}   
\centering
\caption{Difference between our 9.5PN results on $q(u_p)$ and a sample of numerical values from Ref. \cite{Akcay:2015pjz}.}
\begin{ruledtabular}
\begin{tabular}{lllll}
$p=1/u_p$ & $q^{\rm num} -q^{9.5\rm PN}$ & $\Delta q^{\rm num}$ & $\Delta {q}^{9.5\rm PN}$ &  $(q^{\rm num} -q^{9.5\rm PN})/{\rm sup}( \Delta {q})$ \cr
\hline
\hline
6.25&   0.2163&     0.86$\times 10^{-2}$&                                 2.50& 0.087 \cr 
7.5&    0.2821$\times 10^{-1}$& 0.40$\times 10^{-3}$&                     0.53&    0.053\cr 
9.375&  0.1227$\times 10^{-2}$& 0.10$\times 10^{-4}$&                    0.80$\times 10^{-1}$& 0.015\cr 
12&    -0.2334$\times 10^{-3}$& 3.8$\times 10^{-7}$&        0.97$\times 10^{-2}$& -0.024\cr 
15&    -0.8336$\times 10^{-4}$& 3.0$\times 10^{-8}$&        0.14$\times 10^{-2}$& -0.059\cr 
20&    -0.1192$\times 10^{-4}$& 3.4$\times 10^{-8}$&        0.12$\times 10^{-3}$&  -0.10\cr 
25&    -0.2323$\times 10^{-5}$& 1.5$\times 10^{-8}$&        0.17$\times 10^{-4}$& -0.14\cr 
30&    -0.6039$\times 10^{-6}$& 8.2$\times 10^{-9}$&        0.35$\times 10^{-5}$& -0.17\cr 
40&    -0.5814$\times 10^{-7}$& 8.1$\times 10^{-10}$&       2.87$\times 10^{-7}$& -0.20\cr 
50&     0.2448$\times 10^{-8}$& 8.3$\times 10^{-10}$&       4.1$\times 10^{-8}$& 0.060\cr 
60&     0.8505$\times 10^{-8}$& 6.0$\times 10^{-10}$&       8.3$\times 10^{-9}$& 1.0\cr 
75&     0.6899$\times 10^{-8}$& 5.2$\times 10^{-10}$&      1.2$\times 10^{-9}$& 5.9\cr 
100&   -0.4222$\times 10^{-8}$& 4.2$\times 10^{-10}$&      9.4$\times 10^{-11}$& -10.05${}^*$\cr 
200&   -0.4066$\times 10^{-10}$& 2.6$\times 10^{-10}$&     2.1$\times 10^{-13}$& -0.16${}^*$\cr 
400&   -0.4072$\times 10^{-10}$& 1.8$\times 10^{-10}$&     4.7$\times 10^{-16}$& -0.23${}^*$\cr
\end{tabular}
\end{ruledtabular}
\label{tab:4}
\end{table}

\begin{table}
\centering
\caption{Difference between our 9.5PN results on $\rho(u_p)$ and the numerical values given in Ref. \cite{Barack:2010ny}.} 
\begin{ruledtabular}
\begin{tabular}{lllll}
$p=1/u_p$ & $\rho^{\rm num} -\rho^{9.5\rm PN}$ & $\Delta \rho^{\rm num}$ & $\Delta {\rho}^{9.5\rm PN}$ &  $(\rho^{\rm num} -\rho^{9.5\rm PN})/{\rm sup}( \Delta {\rho})$ \cr
\hline
     80& 0.1604$\times 10^{-6}$& 9$\times 10^{-7}$&       4.8$\times 10^{-12}$&       0.18${}^*$\cr 
57.142& -0.4533$\times 10^{-7}$& 6$\times 10^{-7}$&       1.3$\times 10^{-10}$&       -0.076${}^*$\cr 
50&      0.2676$\times 10^{-6}$& 2$\times 10^{-6}$&       4.7$\times 10^{-10}$&       0.13${}^*$\cr 
44.444&  0.6668$\times 10^{-6}$& 2$\times 10^{-6}$&       1.5$\times 10^{-9}$&        0.33${}^*$\cr 
40&      0.6832$\times 10^{-6}$& $10^{-6}$&               4.1$\times 10^{-9}$&        0.68${}^*$\cr 
36.363&  0.5630$\times 10^{-6}$& 8$\times 10^{-7}$&       1.0$\times 10^{-8}$&        0.70${}^*$\cr 
34.2857& 0.1104$\times 10^{-5}$& 8$\times 10^{-7}$&       1.8$\times 10^{-8}$&       1.4${}^*$\cr 
30&      0.1759$\times 10^{-5}$& 4$\times 10^{-7}$&       6.7$\times 10^{-8}$&      4.4${}^*$\cr 
25&      0.2671$\times 10^{-5}$& 3$\times 10^{-7}$&       3.9$\times 10^{-7}$&        6.8\cr 
20&      0.4673$\times 10^{-5}$& 5$\times 10^{-7}$&       3.4$\times 10^{-6}$&        1.4\cr 
19&      0.5111$\times 10^{-5}$& 3$\times 10^{-6}$&       5.6$\times 10^{-6}$&       0.92\cr 
18&      0.5584$\times 10^{-5}$& 2$\times 10^{-6}$&       9.4$\times 10^{-6}$&       0.60\cr 
17&      0.6312$\times 10^{-5}$& 4$\times 10^{-6}$&       1.6$\times 10^{-5}$&        0.39\cr 
16&      0.6621$\times 10^{-5}$& 2$\times 10^{-6}$&       2.9$\times 10^{-5}$&        0.23\cr 
15&      0.8437$\times 10^{-5}$& 3$\times 10^{-6}$&       5.4$\times 10^{-5}$&        0.16\cr 
14&      0.8510$\times 10^{-5}$& 2$\times 10^{-6}$&       1.1$\times 10^{-4}$&        0.081\cr 
13&      0.7202$\times 10^{-5}$& 3$\times 10^{-6}$&       2.1$\times 10^{-4}$&        0.034\cr 
12&      0.4478$\times 10^{-5}$& 3$\times 10^{-6}$&       4.6$\times 10^{-3}$&        0.0097\cr 
11&      0.6084$\times 10^{-6}$& 3$\times 10^{-6}$&       1.1$\times 10^{-3}$&        0.00057\cr 
10&      0.7871$\times 10^{-5}$& 4$\times 10^{-6}$&       2.6$\times 10^{-3}$&        0.0030\cr 
9&       0.1026$\times 10^{-3}$& 5$\times 10^{-6}$&       7.2$\times 10^{-3}$&        0.014\cr 
8.5&     0.2515$\times 10^{-3}$& 6$\times 10^{-6}$&       1.2$\times 10^{-2}$&        0.020\cr 
8&       0.6109$\times 10^{-3}$& 6$\times 10^{-6}$&       2.2$\times 10^{-2}$&        0.028\cr 
7.5&     0.1456$\times 10^{-2}$& $ 10^{-5}$&              4.1$\times 10^{-2}$&        0.036\cr 
7.4&     0.1719$\times 10^{-2}$& 7$\times 10^{-6}$&       4.6$\times 10^{-2}$&        0.037\cr 
7&       0.3543$\times 10^{-2}$& $ 10^{-5}$&              7.8$\times 10^{-2}$&        0.045\cr 
6.8&     0.5096$\times 10^{-2}$& 9$\times 10^{-6}$&       1.0$\times 10^{-1}$&        0.049\cr 
6.5&     0.8825$\times 10^{-2}$& $ 10^{-5}$&              1.6$\times 10^{-1}$&        0.056\cr 
6&       0.2284$\times 10^{-1}$& 4$\times 10^{-5}$&       3.4$\times 10^{-1}$&        0.068\cr
\end{tabular}
\end{ruledtabular}
\label{tab:5}
\end{table}

\begin{figure}
\[
\begin{array}{cc}
\includegraphics[scale=.35]{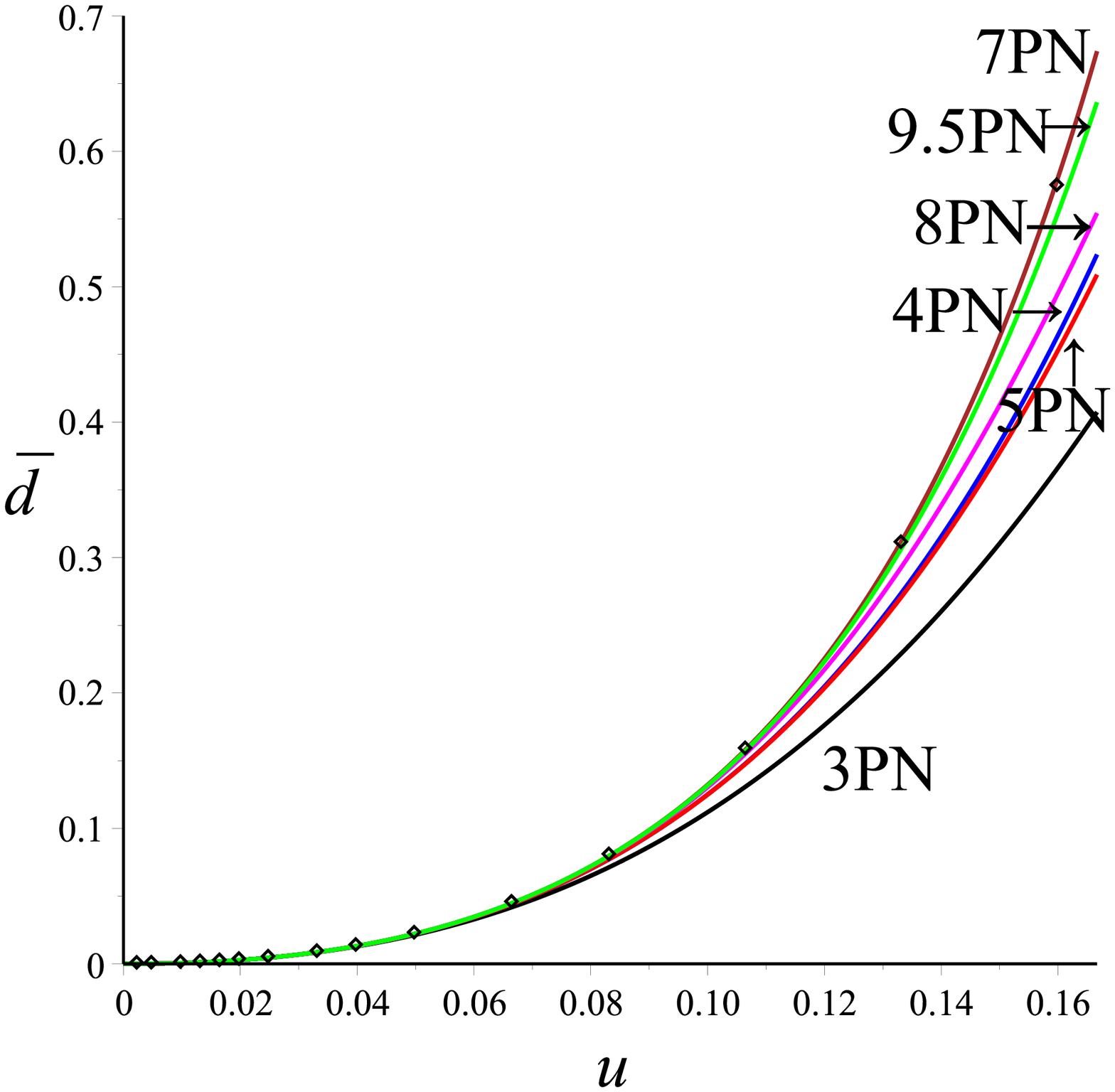}&\qquad
\includegraphics[scale=.35]{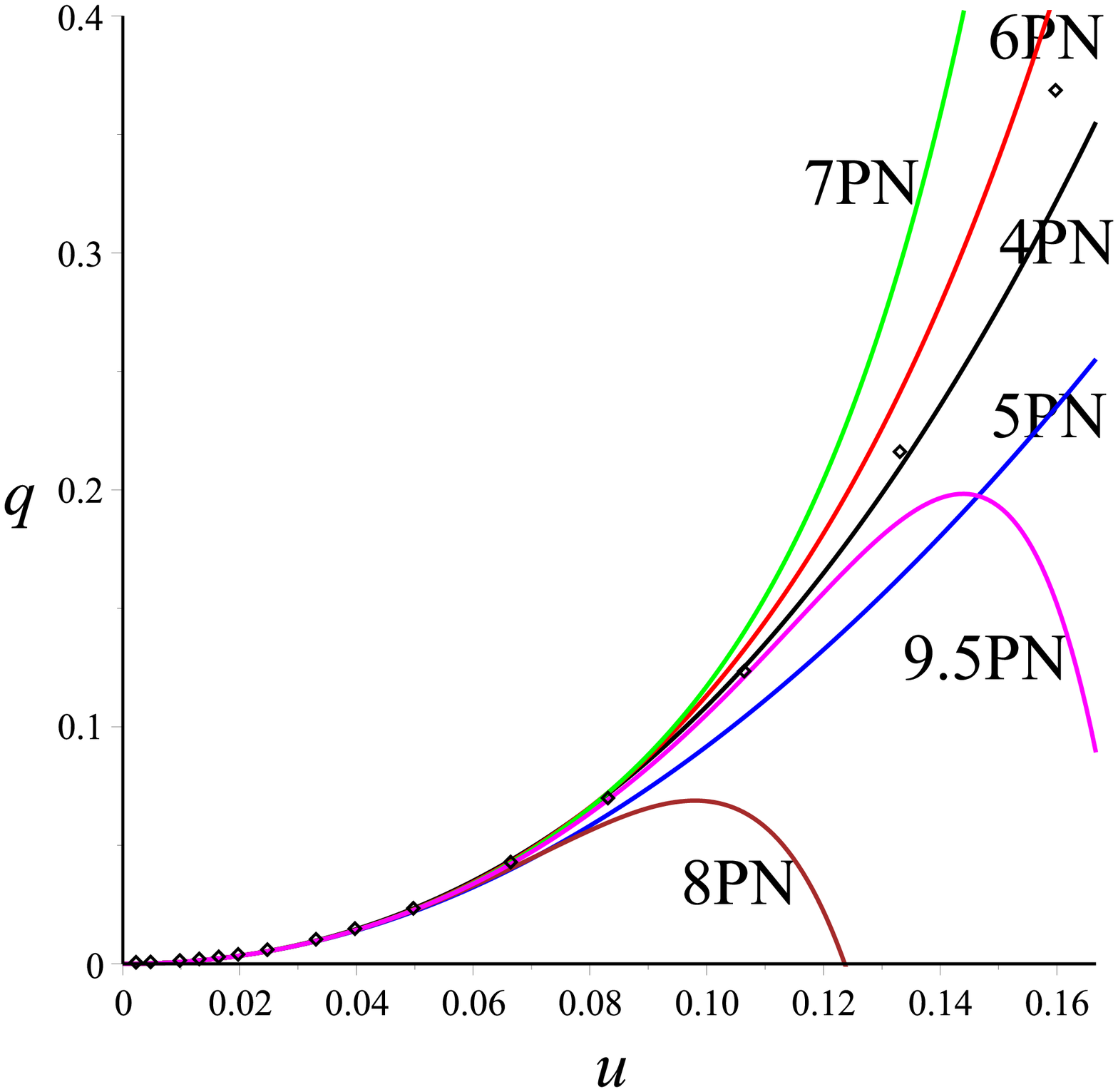}\cr
(a)&\qquad(b)\cr
\end{array}
\]
\caption{
The behavior of various PN-approximants to the EOB functions $\bar d$ and $q$ is shown in panels (a) and (b), respectively,
and compared to a sample of numerical data points from \cite{Akcay:2015pjz}.
\label{fig:1}
}
\end{figure}

\begin{figure}
\[
\begin{array}{cc}
\includegraphics[scale=.35]{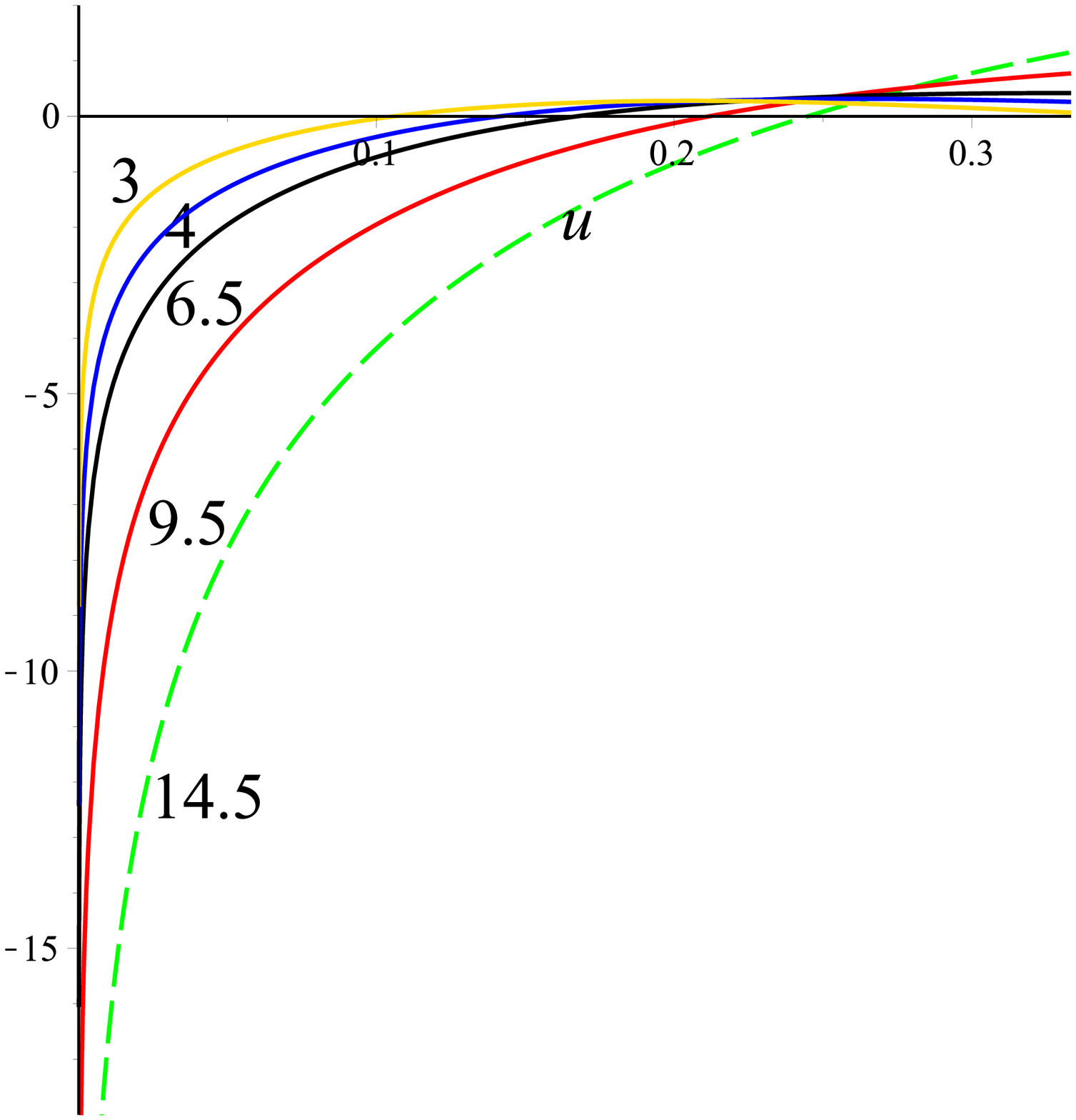}&\qquad
\includegraphics[scale=.35]{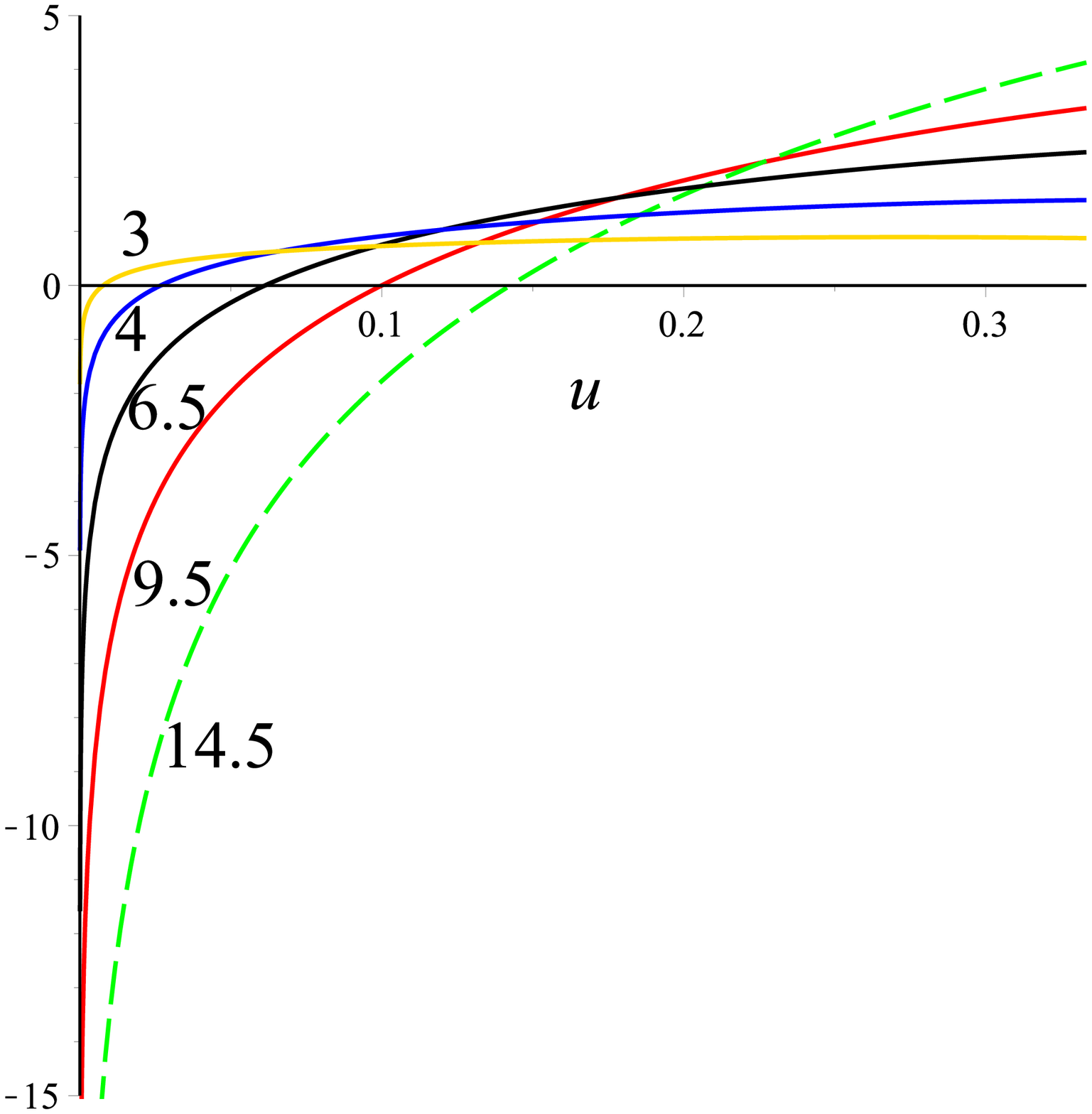}\cr
(a)&\qquad(b)\cr
\end{array}
\]
\caption{
The (base-10) logarithms of twice the fractional theoretical PN errors on $\bar d$ (panel a), and $q$ (panel b),
for $N=[3,4,6.5,9.5,14.5]$ (see text for details).
\label{fig:2}
}
\end{figure}

\end{widetext}

To complete these numerical comparisons by a visual study of the convergence of the PN approximants, we display
in Fig. \ref{fig:1} several successive PN-approximants to the two EOB potentials $\bar d$ and $q$, as well as 
a sample of numerical data points from  \cite{Akcay:2015pjz}.
There is a visible difference between the behavior of the sequence of PN approximants to $\bar d$ and to $q$:
while the successive PN approximants to $\bar d$ seem to exhibit the usual erratic, non-monotonic
\lq\lq convergence" toward the exact (numerical) result, the successive PN approximants to $q$ seem to lose
any \lq\lq convergence" beyond $u \sim 0.12$. [Note, however, that some PN approximants are accidentally closer
to the numerical results than the other ones: especially, the 7PN approximant for $\bar d$ and
the 4PN one for $q$.]

The origin of the latter behavior is rooted in the presence of the
relatively large additional power-law correction $\propto N^4$ in the rescaled PN coefficients
$ {\widehat q}_N \equiv q_N/3^N$, see Eq. \eqref{qN1}. To better study the influence of these power-law corrections to
the basic exponential growth $\propto 3^N$, we plot in Figs. \ref{fig:2} (a) and (b), respectively, 
the {\it fractional} PN errors $\Delta \bar d^ {N \rm PN}/\bar d$, $\Delta q^{ N \rm PN}/q$ \footnote{We use as denominators here the accidentally best PN approximants, i.e. 
$\bar d^{7\rm PN}$ and $q^{4\rm PN}$.}, 
associated with 
the PN remainders at order $N$ PN,
for the PN orders $N= 3,4, 6.5, 9.5$ and $14.5$. Here, the cases $N=3,4$ illustrate the currently fully known
PN knowledge, the case $N=6.5$ illustrates the level of SF PN knowledge \cite{Bini:2015bfb}  for $\bar d$ before the 
present work, $N=9.5$ illustrates the new knowledge brought by the present work, while, finally, $N=14.5$
is included (dashed curve) to illustrate what improvement might bring a much more accurate analytical SF computation of
$\bar d$ and  $q$. For clarity, the vertical axis of these figures plot the base-10 logarithm of {\it twice}
the ratios $\Delta \bar d^ {N PN}/\bar d$, $\Delta q^{ N PN}/q$, so that the crossing of the horizontal axis represents the
location on the $u$ axis where the expected PN  error represents about a $  50\%$ correction to the exact answer
($\Delta f/f=\frac12$).
We can then consider that, from the practical point of view, the crossing of the horizontal axis defines the
right boundary of the domain of validity of the corresponding PN approximant. For instance, we see on 
Fig. \ref{fig:2} (a) that our current 9.5PN approximant to $\bar d$ loses its validity beyond $u \simeq 0.20$,
while Fig. \ref{fig:2} (b) shows that our current 9.5PN approximant to q loses its validity beyond 
the significantly smaller value $u \simeq 0.10$. We note also that an even much improved 14.5PN analytical
knowledge of $q$ would only displace the right boundary of the so-defined domain of validity to $ u \approx 0.15$.
This behavior makes it clear that high-order PN approximants lose any practical interest for representing
the strong-field behavior of the EOB potential $q(u)$. On the other hand, as the 4PN approximant to $q$
is accidentally better than the other ones, it might serve (together with some extra Pad\'e-like factor) 
as a basis for writing an accurate global analytical representation of $q$. Similarly for $\bar d$ when
using the 7PN approximant as a basis. Note, however, that it is urgently needed to go beyond the last stable orbit barrier
at $ u=\frac16$. As the current, precession-based or redshift-based methods are essentially limited
to the range $0<u<\frac16$, it would be interesting, as was emphasized early on \cite{Damour:2009sm},
to use hyperbolic scattering SF computations to explore the EOB potentials in a larger domain of variation.

\section{Analytical $4$PN results for $\delta U^{e^n}$ up to $n=20$}

Ref. \cite{Damour:2015isa} has shown how to convert the nonlocal interaction appearing at the 4PN order \cite{Blanchet:1987wq,Damour:2014jta}
into a specific action-angle Hamiltonian. Moreover,  Ref. \cite{Damour:2015isa} showed also how the latter action-angle Hamiltonian could be formally re-expressed
in terms  of an usual Hamiltonian involving an infinite series of even powers of the radial momentum $p_r$ of the type 
\beq
\label{Q}
\widehat Q(r,p_r) = \sum_{n \geq 2} Q_{2 n}(u;\nu) p_r^{2 n}  
\eeq
with a 4PN value of the $Q$ potential\footnote{The $Q$ potential corresponds to a term in the effective EOB Hamiltonian, following the standard EOB notation}, 
of the type
\begin{eqnarray*}
Q_{2 n}^{4PN}(u;\nu) &=&  u^{5-n} \left( \nu q_{2n}^c + \nu^2  (q'_{4} \delta_n^2+  q'_{6} \delta_n^3)\right.\nonumber\\
&& \left. + \nu^3 (q''_{4} \delta_n^2+  q''_{6} \delta_n^3) \right)\,.
\end{eqnarray*}
In this expression the contributions that are nonlinear in $\nu$ occur only for $n=2$ ($p_r^4$) and $n=3$   ($p_r^6$), and are only contributed by the {\it local}
piece of the Hamiltonian. The {\it nonlocal} piece of the 4PN Hamiltonian only contributes terms linear in $\nu$, which correspond to the 1SF order.
By contrast to the other (locally generated) terms, we see that the 1SF 4PN dynamics contains an infinite number of contributions $\sim \sum_n \nu q^c_{2n}  u^{5-n}  p_r^{2n} $.

Ref.  \cite{Damour:2015isa} has computed the explicit values of the 4PN coefficients $q_4^c, q'_4, q''_4$ and   $q_6^c, q'_6, q''_6$ , and provided  general formulas
for computing the higher-order coefficients $ q_{2n}^c $. Ref. \cite{Hopper:2015icj} has computed the next two (1SF) terms, i.e. $q_8^c$ and $q_{10}^c$,  by another route.
Namely, they  computed by SF methods the 4PN-level contributions to $\delta U^{e^8}$ and $\delta U^{e^{10}}$, and then used the results of \cite{Tiec:2015cxa} to convert
these contributions in terms of the corresponding 1SF, 4PN $Q$-potential contributions  $q_8^c$ and $q_{10}^c$. 
We have verified that the $Q$ results of  Ref. \cite{Hopper:2015icj} , namely
\begin{eqnarray}
q_8^c&=&  -\frac{35772}{175}+\frac{21668992}{45}\ln (2)\nonumber\\
&&+\frac{6591861}{350}\ln (3)-\frac{27734375}{126}\ln (5) \,  , \nonumber\\
q_{10}^c&=& -\frac{231782}{1575}-\frac{408889317632}{212625}\ln (2) \nonumber\\
&&-\frac{22187736351}{28000}\ln (3) +\frac{7835546875}{7776}\ln (5)\nonumber\\
&& +\frac{96889010407}{324000}\ln (7)\, ,
\end{eqnarray}
do agree with the results obtained by the general formulas in   \cite{Damour:2015isa}. 

Using Eqs. (7.5) and (7.7) in   \cite{Damour:2015isa}  we have computed the coefficients  $q^c_{2n}$ for $n$ varying between $6$ and $10$. Our results read
\begin{widetext}
\begin{eqnarray}
q_{12}^c &=& -\frac{252412}{2475} -\frac{71310546875}{24948} \ln(5)+\frac{163796987511}{38500} \ln(3)-\frac{96889010407}{29700} \ln(7)+\frac{7057329658112}{779625} \ln(2)\nonumber\\
&\approx & 0.0018257727627680511315034555270224294943015  \,  ,  \nonumber\\
& \nonumber\\
q_{14}^c &=& -\frac{10281865679266304}{212837625} \ln(2)-\frac{877810440113163}{112112000} \ln(3)+\frac{11259375010387063}{667180800} \ln(7)-\frac{3079166}{45045}\nonumber\\
&&+\frac{1878421041015625}{326918592} \ln(5)\nonumber\\
& \approx &   0.0003789439085938192497702387294880213660917 \,  ,  \nonumber\\
q_{16}^c &=&\frac{417442117895446016}{1915538625}\ln(2)-\frac{827476034230539}{22422400}\ln(3)-\frac{401306}{9009}-\frac{2793081608858259887}{50038560000}\ln(7)\nonumber\\
&& -\frac{1153146534765625}{980755776}\ln(5)\nonumber\\
& \approx &     -0.000094237712462701263218285845763487007072  \,  ,  \nonumber\\
q_{18}^c &=&   -\frac{22451335308782004224}{28733079375}\ln(2)+\frac{7400249944258160101211}{363771233280000}\ln(11)-\frac{184181968578981640625}{2134124568576}\ln(5)\nonumber\\
&&-\frac{1409537}{49725}+\frac{993339887626452455369}{7423902720000}\ln(7)+\frac{303173836939989783}{896896000}\ln(3)\nonumber\\
& \approx &    -0.000012390367224863884457529152672385306211 \,  ,  \nonumber\\
q_{20}^c &=& -\frac{12248956}{692835}-\frac{939831742920786332853797}{3455826716160000}\ln(11)-\frac{967033461070767993388297}{3878989171200000}\ln(7)\nonumber\\
&& +\frac{47970130001158879756288}{19887395653125}\ln(2)-\frac{4894596811860934937067}{3621217600000}\ln(3)+\frac{107731911758417652734375}{182467650613248}\ln(5)\nonumber\\
& \approx &  0.00000218442987916976096395571442115985913\,.
\end{eqnarray}
 Finally,  following the procedure outlined in Ref. \cite{Tiec:2015cxa}, it is straightforward to determine from the so-determined 4PN-accurate EOB $Q$-potential
  the corresponding 4PN-accurate expression of the redshift 
 coefficient functions $\delta U^{e^n}(u_p)$. We found for $n= 6, 8, 10 $ (already given in   \cite{Hopper:2015icj}) and $n=12, 14, 16, 18, 20$ (new results):

\begin{eqnarray}
\delta U^{e^6}&=&-\frac52  u_p^3+\left(-\frac{475}{12}+\frac{41}{128}\pi^2\right) u_p^4+\left(-\frac{52877}{180}+\frac{178288}{5}\ln(2)-\frac{1994301}{160}\ln(3)-\frac{1953125}{288}\ln(5)-16\gamma\right. \nonumber\\
&& \left. +\frac{3385}{4096}\pi^2-8\ln(u_p)\right)u_p^5\,,   \nonumber\\
\delta U^{e^8}&=& \frac{15}{64}u_p^3+\left(-\frac{1171}{384}+\frac{287}{4096}\pi^2  \right)u_p^4+\left( -\frac{55}{12}\ln u_p -\frac{24619}{384}-\frac{55}{6}\gamma +\frac{327115}{196608}\pi^2 -\frac{15967961}{90}\ln (2)\right. \nonumber\\
&&\left.  +\frac{11332791}{1280}\ln (3) +\frac{162109375}{2304}\ln (5) \right)u_p^5\,, \nonumber\\
\delta U^{e^{10}}&=& \frac{3}{64}u_p^3 +\left(-\frac{115}{128}+\frac{123}{4096}\pi^2  \right)u_p^4 +\left( -\frac{329}{240}\ln u_p -\frac{1933}{3840} -\frac{329}{120}\gamma +\frac{172697}{393216}\pi^2 +\frac{18046622551}{27000}\ln (2)\right.\nonumber\\
&&\left.  +\frac{203860829079}{1024000}\ln (3) -\frac{74048828125}{221184}\ln (5)-\frac{678223072849}{9216000}\ln (7)\right)u_p^5\,,
\end{eqnarray}

and

\begin{eqnarray}
\delta U^{e^{12}}&=&\frac{5}{512} u_p^3 +\left(-\frac{1909}{3072}+\frac{533}{32768}\pi^2\right) u_p^4\nonumber\\
&& +  \left(-\frac{104557}{46080}+\frac{1655}{8192}\pi^2-\frac{95932245107}{36000}\ln(2)-\frac{211}{320}\ln(u_p)-\frac{4936871473659}{4096000}\ln(3)+\frac{678223072849}{819200}\ln(7)\right.\nonumber\\
&&\left.
-\frac{211}{160}\gamma+\frac{285888671875}{294912}\ln(5)\right) u_p^5\,,\nonumber\\
\delta U^{e^{14}}&=& \left(-\frac{5}{12}+\frac{41}{4096}\pi^2\right) u_p^4\nonumber\\
&&+  \left(-\frac{135071}{107520}+\frac{4311583788974229}{1605632000}\ln(3)-\frac{2758333237276883}{637009920}\ln(7)+\frac{195921190766921}{15876000}\ln(2)\right.\nonumber\\
&&\left.
-\frac{2984729833984375}{1560674304}\ln(5) +\frac{89395}{786432}\pi^2-\frac{73}{192}\ln(u_p)-\frac{73}{96}\gamma  \right) u_p^5\,,  \nonumber\\
\delta U^{e^{16}}&=&-\frac{45}{16384} u_p^3+\left(\frac{7011}{1048576}\pi^2-\frac{9525}{32768}\right) u_p^4  \nonumber\\
&&+ \left(\frac{3606265}{50331648}\pi^2-\frac{187}{384}\gamma-\frac{866799}{1146880}+\frac{151266508326642969}{25690112000}\ln(3)+\frac{30647775337890625}{24970788864}\ln(5)\right.\nonumber\\
&&\left.
-\frac{187}{768}\ln(u_p)+\frac{132130740829369331}{9437184000}\ln(7) -\frac{9809130397488463}{190512000}\ln(2)  \right) u_p^5\,,
\nonumber\\
\delta U^{e^{18}}&=&-\frac{55}{16384}u_p^3 +\left(-\frac{20755}{98304}+\frac{4961}{1048576}\pi^2\right) u_p^4\nonumber\\
&& + \left(-\frac{1985885}{4128768}+\frac{1632805}{33554432}\pi^2-\frac{685}{4096}\ln(u_p)-\frac{81402749386839761113321}{21574761578496000}\ln(11)-\frac{685}{2048}\gamma\right.\nonumber\\
&&\left.
+\frac{926296539361158203125}{57532697542656}\ln(5)-\frac{13828959005709035994883}{440301256704000}\ln(7)\right. \nonumber\\
&& \left. -\frac{27896787814453074891}{411041792000}\ln(3)+\frac{14107331956051038263}{82301184000}\ln(2)\right) u_p^5\,, \nonumber\\
\delta U^{e^{20}}&=& -\frac{429}{131072}u_p^3+\left(-\frac{124795}{786432}+\frac{29315}{8388608}\pi^2\right) u_p^4\nonumber\\ 
&& +\left(-\frac{395456141}{1238630400}-\frac{240907615282410313097}{503884800000}\ln(2)-\frac{29689}{245760}\ln(u_p)-\frac{230415740222184068359375}{2071177111535616}\ln(5)\right. \nonumber\\
&& +\frac{1551270323409360145587}{5872025600000}\ln(3)+\frac{2290704611907930887405849}{44030125670400000}\ln(7)+\frac{21408923088738857172803423}{431495231569920000}\ln(11)\nonumber\\
&& \left.+\frac{7017461}{201326592}\pi^2-\frac{29689}{122880}\gamma\right)u_p^5\,.   
\end{eqnarray}

The conversion of these results in terms of the $e^n$ expansion of $\delta z_1= - \delta U/U_0^2$ reads

\begin{eqnarray}
\delta z_1^{e^{12}}&=&   \frac{25}{512} u_p^3+\left(-\frac{533}{32768}\pi^2+\frac{3391}{3072}\right)u_p^4\nonumber\\
&+&\left(-\frac{7973}{32768}\pi^2+\frac{95932245107}{36000}\ln(2)+\frac{211}{320}\ln(u_p)+\frac{204491}{23040}+\frac{4936871473659}{4096000}\ln(3)-\frac{678223072849}{819200}\ln(7)\right.\nonumber\\
&&\left.
+\frac{211}{160}\gamma-\frac{285888671875}{294912}\ln(5)\right) u_p^5\,, \nonumber\\
\delta z_1^{e^{14}}&=&   \frac{15}{512}u_p^3+\left(\frac{485}{768}-\frac{41}{4096}\pi^2\right)u_p^4\nonumber\\
&+&\left(-\frac{4311583788974229}{1605632000}\ln(3)+\frac{2758333237276883}{637009920}\ln(7)+\frac{104249}{26880}-\frac{195921190766921}{15876000}\ln(2)\right.\nonumber\\
&&\left.
+\frac{2984729833984375}{1560674304}\ln(5)-\frac{104155}{786432}\pi^2+\frac{73}{192}\ln(u_p)+\frac{73}{96}\gamma\right) u_p^5\,,\nonumber\\
\delta z_1^{e^{16}}&=&   \frac{315}{16384} u_p^3+\left(-\frac{7011}{1048576}\pi^2+\frac{13215}{32768}\right) u_p^4\nonumber\\
&+&\left(-\frac{4108105}{50331648}\pi^2+\frac{187}{384}\gamma+\frac{586431}{286720}-\frac{151266508326642969}{25690112000}\ln(3)-\frac{30647775337890625}{24970788864}\ln(5)+\frac{187}{768}\ln(u_p)\right. \nonumber\\
&& \left. -\frac{132130740829369331}{9437184000}\ln(7)+\frac{9809130397488463}{190512000}\ln(2)\right) u_p^5\,, 
\nonumber\\
\delta z_1^{e^{18}}&=&   \frac{55}{4096} u_p^3+\left(-\frac{4961}{1048576}\pi^2+\frac{27205}{98304}\right) u_p^4\nonumber\\
&&+\left(\frac{685}{4096}\ln(u_p)-\frac{1829605}{33554432}\pi^2+\frac{81402749386839761113321}{21574761578496000}\ln(11)+\frac{4960745}{4128768}+\frac{685}{2048}\gamma\right. \nonumber\\
&&-\frac{926296539361158203125}{57532697542656}\ln(5)+\frac{13828959005709035994883}{440301256704000}\ln(7)+\frac{27896787814453074891}{411041792000}\ln(3)\nonumber\\
&& \left.-\frac{14107331956051038263}{82301184000}\ln(2)\right) u_p^5\,,  \nonumber\\
\delta z_1^{e^{20}}&=&  \frac{1287}{131072} u_p^3+\left(\frac{157201}{786432}-\frac{29315}{8388608}\pi^2  \right) u_p^4\nonumber\\
&& +\left(\frac{234834179}{309657600}+\frac{240907615282410313097}{503884800000}\ln(2)+\frac{29689}{245760}\ln(u_p)+\frac{230415740222184068359375}{2071177111535616}\ln(5)\right. \nonumber\\
&&-\frac{1551270323409360145587}{5872025600000}\ln(3)-\frac{2290704611907930887405849}{44030125670400000}\ln(7)-\frac{21408923088738857172803423}{431495231569920000}\ln(11)\nonumber\\
&&\left. -\frac{7764317}{201326592}\pi^2+\frac{29689}{122880}\gamma\right) u_p^5\,.   
\end{eqnarray}

\end{widetext}

\section{Discussion}

We have improved the analytical knowledge of the eccentricity-expansion of the Detweiler-Barack-Sago redshift invariant (in a Schwarzschild spacetime) in several different ways. First, we have analytically computed the $e^2$ and $e^4$ contributions
to the 1SF contribution to the average redshift up to the 9.5th post-Newtonian order (included). For the $e^2$
contribution this is an improvement by three PN orders compared to previous knowledge. For the $e^4$ contribution this
is an improvement by five-and-a-half PN orders compared to previous knowledge. We have also provided for the first
time the $e^{12},  e^{14}, e^{16}, e^{18}$, and $e^{20}$ contributions to the 4PN approximation. We have then converted this new analytical information in terms of corresponding dynamically relevant effective-one-body (EOB)
 potentials: $\bar d(u)$, $\rho(u)$ and $q(u)$.

We have shown how to estimate the order of magnitude of the  coefficients of the PN expansions of the EOB potentials
$a(u)$,  $\rho(u)$, $\bar d(u)$, and $q(u)$. See Eqs. \eqref{aN}, \eqref{rhoN1new}, \eqref{rhoN2}, \eqref{dN1}, \eqref{dN2},
\eqref{qN1}, \eqref{qN2}. We then used this knowledge to estimate the remainder terms in our current 9.5PN-accurate
expansions. Let us note that it would be interesting to refine our estimates, and, in particular, to numerically study the behavior
of the rescaled coefficients ${\widehat a}_N \equiv a_N/3^N$ for very large values of $N$. We gave arguments suggesting
a slow decrease ${\widehat a}_N \sim N^{-\frac12}$, but found no evidence for it up to $N = 23$.
This might be due to a transient behavior proportional to $ (1-3u)^{-1}= \sum_N (3 u)^N$ of $a(u)$ before it zooms on its near-lightring behavior
$\propto (1-3u)^{-\frac12}$. [Indeed, Ref. \cite{Akcay:2012ea} found that the rescaled function $\hat a_E(u)$ increases
rather steeply (from 1 to $\simeq 10$) as $u$ varies between 0 and $\frac13$. This increase is roughly proportional (modulo
an essentially linear function) to $\propto (1-3u)^{-\frac12}$ (before levelling off) and might explain the transient appearance of a growth
of $a(u)$ roughly proportional to $ (1-3u)^{-\frac12} E(u)= (1-2u) (1-3u)^{-1}$.]

We compared our 9.5PN-accurate analytical representations of the functions $\rho(u)$, $\bar d(u)$, and $q(u)$ to the
currently published numerical SF evaluations of these functions \cite{Barack:2010ny,Akcay:2015pjz}. The
results of our comparisons are given in Tables \ref{tab:3}, \ref{tab:4} and \ref{tab:5}.  The analytical/numerical agreement
is  fully satisfactory, in view of the estimated theoretical and numerical errors. [It suggests that both types of errors
have been often slightly overestimated.]

We studied the convergence of the successive PN approximants to both $\bar d(u)$ and $q(u)$, see the two
panels of Fig. \ref{fig:1}. 
The newest result of this study is the particularly unsatisfactory convergence, near $ u \sim 0.1$, of the successive
PN approximants to $q(u)$. We explained this worst-than-usual behavior of PN approximants by the presence of
a large power-law subleading correction $\propto N^4$ to the exponential growth $\propto 3^N$ of the PN coefficients
of $q(u)$. This $N^4$ factor underlies the poor accuracy of the 9.5PN approximant in the relatively weak-field
domain $ u \sim 0.1$. See second panel of Fig. \ref{fig:2}. We leave to future work the 
construction of accurate hybrid PN-SF analytical
representations of the EOB potentials $\bar d(u)$, and $q(u)$, valid both in the weak-field and the strong-field
domains. 

We emphasized that  a comparison between our 9.5PN analytical 
computation of the precession function $\rho(u)$ (which combines SF theory with the eccentric first law \cite{Tiec:2015cxa}) and of  high-accuracy SF numerical computations
of the purely dynamical precession of eccentric orbits (as in \cite{Barack:2010ny}) would be a useful check of the assumptions underlying the theoretical bridges (EOB and the first law of binary mechanics) which have been recently
quite useful for connecting SF and PN results. 

We recalled that the  use of precession-based or redshift-based SF methods currently limit
the computation of the EOB potentials $\bar d(u)$ and $q(u)$ to the medium-strong-field domain
$0<u<\frac16$. It would be interesting, as was pointed out in \cite{Damour:2009sm},
to use hyperbolic scattering SF computations to explore the EOB potentials in a larger domain of variation.
In particular, one would like to confirm the conjecture \cite{Akcay:2012ea} that
$\bar d(u)$ and $\rho(u)$ [which are both related to $a''(u)$] diverge  (when $u \to \frac13$) like $(1-3u)^{-\frac52}$.
One similarly expects that $q(u)$ [related to $a''''(u)$] will diverge like $(1-3u)^{-\frac92}$. The power-law growths found
in the rescaled PN coefficients $ \widehat{\bar d}_N, \widehat \rho_N, \widehat q_N$ suggest such strong lightring singularities
(though there is a mismatch of a missing factor $N^{-\frac12}$ in the observed growths).

\appendix

\section{Combined use of RWZ approach, PN solutions and MST technique}
\label{appRWZ}

Our analytical computation of the conservative SF effects along an eccentric orbit in a Schwarzschild background follows the approach originally introduced  in Ref. \cite{Bini:2013zaa} and then standardized in a sequence of successive works \cite{Bini:2013rfa,Bini:2014ica,Bini:2014zxa,Bini:2015bla,Bini:2015mza,Kavanagh:2015lva}.  
The main steps (together with some of the most important computational details) are listed below [see our previous papers for the notation, which we follow here.]

The Detweiler-Barack-Sago \cite{Detweiler:2008ft,Barack:2011ed} inverse redshift invariant function for eccentric orbits is defined as
\beq
\label{I1}
U\left(m_2\Omega_r, m_2\Omega_\phi, \frac{m_1}{m_2}\right)= \frac{\displaystyle\oint dt}{\displaystyle\oint d\tau}=\frac{T_r}{{\mathcal T}_r}\,,
\eeq
where all quantities refer to the perturbed spacetime metric (see Eq. \eqref{gperturbed} below).
The symbol $\oint $ denotes an integral over a radial period (from periastron to periastron) so that $T_r=\oint dt$ denotes the coordinate-time period and ${\mathcal T}_r=\oint d\tau$ the proper-time period. 
The first-order SF contribution $\delta U$ to the function \eqref{I1}, defined by 
\begin{eqnarray}
U\left(m_2\Omega_r, m_2\Omega_\phi, \frac{m_1}{m_2}\right)&=&U_0\left(m_2\Omega_r, m_2\Omega_\phi\right)\nonumber\\
&+& \frac{m_1}{m_2}\delta U\left(m_2\Omega_r, m_2\Omega_\phi\right)\nonumber\\
&+& O\left( \frac{m_1^2}{m_2^2} \right)\,,
\end{eqnarray}
represents a gauge-invariant measure of the $O(m_1/m_2)$ conservative SF effect on eccentric orbits.
It is a function of the two fundamental frequencies of the orbit, i.e., the radial frequency $\Omega_r=2\pi/T_r$ and the mean azimuthal frequency $\Omega_\phi=\Phi/T_r$, where $\Phi$ is the angular advance during one radial period $T_r$, and is conveniently expressed in terms of the dimensionless semi-latus rectum $p$ and the eccentricity $e$ of the unperturbed orbit, i.e., $\delta U=\delta U(p,e)$.
It is given in terms of the $O(m_1/m_2)$ metric perturbation $h_{\mu\nu}$, where 
\beq \label{gperturbed}
g_{\mu\nu}(x^\alpha; m_1, m_2)=g^{(0)}_{\mu\nu}(x^\alpha; m_2)+\frac{m_1}{m_2} h_{\mu\nu}(x^\alpha)+O\left( \frac{m_1^2}{m_2^2} \right)\,,
\eeq
[with $g^{(0)}_{\mu\nu}(x^\alpha; m_2)$ being the Schwarzschild metric of mass $m_2$] by the following
 time average
\beq
\label{delta_U1}
\delta U (p,e)=\frac12 \, U_{0}^2\langle h_{uk}\rangle_{t}\,.
\eeq
Here, we have expressed $\delta U$ (which is originally defined as a {\it proper} time $\tau$ average \cite{Barack:2011ed}) in terms of the {\it coordinate} time $t$ average of the mixed contraction $h_{uk}=h_{\mu\nu}u^\mu k^\nu$ where $u^\mu\equiv u^t k^\mu$, $u^t=dt/d\tau$ and $k^\mu\equiv \partial_t +dr/dt\partial_r +d\phi/dt \partial_\phi$. [Note that in the present eccentric case the so-defined $k^\mu=u^\mu/u^t$ is no longer a Killing vector.] In Eq. \eqref{delta_U1} we considered $\delta U$ as a function of the dimensionless semi-latus rectum $p$ and eccentricity  $e$ (in lieu of $m_2\Omega_r$,  $m_2\Omega_\phi$) of the {\it unperturbed} orbit, as is allowed in a first-order SF quantity. In addition, $U_0$ denotes the proper-time average of $u^t=dt/d\tau$ along the unperturbed orbit, i.e., the ratio $U_0={T_r}/{{\mathcal T}_r}|_{\rm unperturbed}$.

The correction $\delta U $ is equivalent to the correction $\delta z_1$ to the (coordinate-time) averaged redshift $z_1$
\beq
z_1=\left\langle \frac{d \tau}{dt}\right\rangle_t = \left(\left\langle \frac{dt}{d \tau}\right\rangle_\tau \right)^{-1}=U^{-1}\,,
\eeq
namely
\beq
\label{eq_z1}
\delta z_1=-\frac{\delta U}{U_0^2}=-\frac12 \langle h_{uk}\rangle_{t}\,.
\eeq

\subsection{Unperturbed particle motion}

Up to order $e^4$ included, the unperturbed {\it eccentric} particle motion  $r_0(t)$, $\phi_0(t)$ is explicitly given by 

\begin{widetext}

\begin{eqnarray}
r_0(t) &=& R_0+e R_1 (\cos(\Omega_{r0} t)-1)+e^2 R_2(\cos(2 \Omega_{r0} t)-1)\nonumber\\
&& +e^3 [R_{3c3}(\cos(3 \Omega_{r0} t)-1)+R_{3c1}(\cos(\Omega_{r0} t)-1)+R_{3s1} t\sin(\Omega_{r0} t)]\nonumber\\
&&+e^4 [R_{4c2}(\cos(2 \Omega_{r0} t)-1)+R_{4s2} t\sin(2 \Omega_{r0} t)+R_{4c4}(\cos(4\Omega_{r0} t)-1)]\,,\nonumber\\
\phi_0(t) &=& \Omega_{\phi 0} t+e \Phi_1\sin(\Omega_{r0} t)+e^2\Phi_2\sin(2 \Omega_{r0} t)\nonumber\\
&&+e^3 [\Phi_{3c1} t\cos(\Omega_{r0} t)+\Phi_{3s1}\sin(\Omega_{r0} t)+\Phi_{3s3}\sin(3 \Omega_{r0} t)]\nonumber\\
&&+e^4 [\Phi_{4c2}t\cos(2 \Omega_{r0} t)+\Phi_{4s2}\sin(2 \Omega_{r0} t)+\Phi_{4s4}\sin(4\Omega_{r0} t)]\,, 
\end{eqnarray}
with
\begin{eqnarray}
R_0 &=& m_2 p\,(1+ e+ e^2+ e^3+ e^4)\,,\nonumber\\ 
R_1 &=& m_2 p\,,\nonumber\\ 
R_2 &=& -m_2 p\,\frac{p^2-11 p+26}{2(p-2)(p-6)}\,,\nonumber\\ 
R_{3c1} &=& m_2 p\,\frac{10 p^4-124 p^3+385 p^2+220 p-1404}{16(p-2)^2 (p-6)^2}\,,\nonumber\\ 
R_{3c3} &=& m_2 p\,\frac{6 p^4-132 p^3+1023 p^2-3292 p+3708}{16(p-2)^2 (p-6)^2}\,,\nonumber\\ 
R_{3s1} &=& m_2\,\frac{3(2 p^3-32 p^2+165 p-266)}{ 4p (p-6)^{3/2}(p-2)}\,,\nonumber\\ 
R_{4c2} &=& -m_2 p\,\frac{p^5-4 p^4-203 p^3+2337 p^2-9192 p+12228}{6(p-2)^2 (p-6)^3}\,,\nonumber\\ 
R_{4c4} &=& -m_2 p\,\frac{168630 p^2-310092 p-46697 p^3+6951 p^4-528 p^5+16 p^6+226968}{48(p-2)^3 (p-6)^3}\,,\nonumber\\ 
R_{4s2} &=& -m_2\,\frac{ 3(2 p^3-32 p^2+165 p-266) (p^2-11 p+26)}{ 4p (p-6)^{5/2}(p-2)^2}\,,
\end{eqnarray}
and
\begin{eqnarray} 
\Phi_1 &=& -2 \frac{(p-3) p^{1/2}}{(p-2) (p-6)^{1/2}}\,,\nonumber\\ 
\Phi_2 &=& \frac14 \frac{ (5 p^3-64 p^2+250 p-300) p^{1/2}}{ (p-6)^{3/2} (p-2)^2}\,,\nonumber\\ 
\Phi_{3c1} &=&\frac32 \frac{ (p-3) (2 p^3-32 p^2+165 p-266)}{p^{3/2}(p-6)^2 (p-2)^2}\,,\nonumber\\ 
\Phi_{3s1} &=& \frac18 \frac{(2 p^5-74 p^4+855 p^3-4261 p^2+9264 p-6948) p^{1/2}}{ (p-6)^{5/2}(p-2)^3}\,,\nonumber\\ 
\Phi_{3s3} &=& -\frac{1}{24}\frac{(26 p^5-594 p^4+5187 p^3-21545 p^2+42480 p-31860) p^{1/2}}{ (p-6)^{5/2}(p-2)^3}\,,\nonumber\\ 
\Phi_{4c2} &=& -\frac{3}{8}\frac{ (5 p^3-64 p^2+250 p-300) (2 p^3-32 p^2+165 p-266)}{p^{3/2} (p-6)^3 (p-2)^3}\,,\nonumber\\ 
\Phi_{4s2} &=& -\frac{1}{48}\frac{(22 p^7-972 p^6+16191 p^5-136892 p^4+644034 p^3-1695084 p^2+2313960 p-1262160)p^{1/2}}{ (p-6)^{7/2}(p-2)^4}\,,\nonumber\\ 
\Phi_{4s4} &=& \frac{1}{192}\frac{ (206 p^7-6804 p^6+93327 p^5-687580 p^4+2932674 p^3-7231980 p^2+9545448 p-5206608)p^{1/2}}{ (p-6)^{7/2}(p-2)^4}\,.
\end{eqnarray}
The $m_2$-adimensionalized orbital frequencies of the radial and azimuthal motions are given by
\begin{eqnarray}
\Omega_{r0}&=& \frac{ (p-6)^{1/2}}{ p^2} -\frac34 \frac{(2 p^3-32 p^2+165 p-266)}{p^2 (p-2) (p-6)^{3/2} } e ^2\nonumber\\
&&+\frac{3}{64} \frac{ (8 p^7-336 p^6+5724 p^5-51456 p^4+263441 p^3-764550 p^2+1152396 p-681224)}{ p^2 (p-6)^{7/2}(p-2)^3 } e^4\,,\nonumber\\
\Omega_{\phi 0} &=& \frac{1}{p^{3/2}}-\frac32 \frac{ (p^2-10 p+22)}{ (p-2) (p-6) p^{3/2}} e^2
+\frac{3}{16}\frac{ (2 p^6-72 p^5+993 p^4-6786 p^3+24250 p^2-42528 p+27864)}{ (p-2)^3 (p-6)^3 p^{3/2}}e^4\,,
\end{eqnarray}
respectively.
Finally, the (unperturbed) redshift variable $U_0=T_{r0}/{\mathcal T}_{r0}$ is given by
\begin{eqnarray} 
\label{U0}
U_0&=&
\frac{p^{1/2}}{(p-3)^{1/2}}
-\frac32\frac{(p^2-10p+22)p^{1/2}}{(p-2) (p-6)(p-3)^{3/2}}e^2\nonumber\\
&&
-\frac38\frac{(p^6-6p^5-163p^4+2188p^3-10565p^2+22860p-18612)p^{1/2}}{(p-2)^3 (p-6)^3(p-3)^{5/2}}e^4
+O(e^5)\,.
\end{eqnarray}

\subsection{Source terms}		
		
We first compute the (nine) source terms in the RWZ perturbation approach, namely
\beq
A_{lm}^{(0)}(t,r)=m_1 u^t \left(\frac{dr_0}{dt}\right)^2\frac{1}{(r-2m_2)^2}\delta (r-r_0(t)) \, e^{-i m \phi_0(t)} Y^{*}_{lm}\left(\frac{\pi}2\right)\,,
\eeq
etc., where   $Y_{lm}(\theta)$ denotes the value of the usual spherical harmonics at $\phi=0$, while $Y'{}_{lm}(\theta)$ denotes its $\theta$-derivative, and consider then their Fourier transform
\beq
A_{lm \omega}^{(0)}(r)=\int_{-\infty}^\infty e^{i\omega t}A_{lm}^{(0)}(t,r) dt\,,
\eeq
etc. The result  (after expanding in powers of the eccentricity through $e^4$) is of the form
\beq
A_{l m  \omega}^{(0)}(r)=\sum_{n=-4}^4 [c_n(r) \delta_n+\tilde c_n(r) \delta'_n]\,,\qquad
c_n(r)= \sum_{k=0}^4 c_{n,k}(r_0)\delta^{(k)} (r-r_0)\,,\qquad
\tilde c_n(r)= \sum_{k=0}^4 d_{n,k}(r_0)\delta^{(k)} (r-r_0)\,,
\eeq
with $r_0=m_2p$ and 
\beq
\delta_{ n}=\delta (\omega -\omega_{m,n})\,,\qquad \omega_{m,n}=m  \Omega_{\phi0} +  n\Omega_{r0}\,,
\eeq
so that $\delta_{n= 0}=\delta(\omega-m \Omega_{\phi0})$, $\delta_{n= +1}=\delta(\omega-m \Omega_{\phi0}-\Omega_{r0})$, etc. 
The various quantities $\delta_{ n}$, $c_n(r)$, etc. also depend on $l,m,\omega$, even if not shown explicitly to ease the notation. 
With the coefficients $A_{l m \omega}^{(0)}(r)$, etc.  one computes the  odd- and even-Zerilli sources.
In order to write a single Regge-Wheeler (RW) equation for both cases, the even-Zerilli sources should be mapped into certain (different) even sources, the associated  map requiring an extra $r$-derivative.

Summarizing, the odd sources are of the form
\begin{eqnarray}
S^{\rm(odd)}_{lm\omega}(r)&=&\sum_{n=-4}^4 s^{\rm(odd)}_n(r) \delta_n+\sum_{n=-2}^2 \tilde s^{\rm(odd)}_n(r) \delta'_n\,,\nonumber\\
s^{\rm(odd)}_n(r)&=& \sum_{k=0}^5 s^{\rm(odd)}_{n,k}(r_0)\delta^{(k)} (r-r_0)\,,\qquad
\tilde s^{\rm(odd)}_n(r)= \sum_{k=0}^3 s^{\rm(odd)}_{n,k}(r_0)\delta^{(k)} (r-r_0)\,,
\end{eqnarray}
while the even sources are of the form
\begin{eqnarray}
S^{\rm(even)}_{lm\omega}(r)&=&\sum_{n=-4}^4 s^{\rm(even)}_n(r) \delta_n+\sum_{n=-2}^2 \tilde s^{\rm(even)}_n(r) \delta'_n\,,\nonumber\\
s^{\rm(even)}_n(r)&=& \sum_{k=0}^6 s^{\rm(even)}_{n,k}(r_0)\delta^{(k)} (r-r_0)\,,\qquad
\tilde s^{\rm(even)}_n(r)= \sum_{k=0}^4 \tilde s^{\rm(even)}_{n,k}(r_0)\delta^{(k)} (r-r_0)\,,
\end{eqnarray}
both of them satisfying the RW equation
\beq
\label{eq:RW}
{\mathcal L}^{(r)}_{\rm (RW)}[R_{lm\omega }^{\rm (even/odd)}]=S_{lm\omega}^{\rm (even/odd)}(r)\,,
\eeq
where ${\mathcal L}^{(r)}_{\rm (RW)}$ denotes the RW operator 
\begin{eqnarray}
\label{eq:operator}
{\mathcal L}^{(r)}_{\rm (RW)}&=& \frac{d^2}{dr_*^2}   +[\omega^2 -V_{\rm (RW)}(r)]\,,
\end{eqnarray}
with $d/dr_* = f(r) d/dr$ (with $f(r)\equiv 1-2m_2/r$), and a RW potential
\begin{eqnarray}
\label{eq:potential}
V_{\rm (RW)}(r)&=&f(r)\left(\frac{l(l+1)}{r^2}-\frac{6 \, m_2}{r^3}  \right)\,.
\end{eqnarray}

\subsection{Green function}

The Green function of the RW equation \eqref{eq:RW} reads as
\beq
G_{lm\omega}(r,r')=\frac{1}{W_{lm\omega}}\left[R_{\rm in}^{lm\omega}(r)R_{\rm up}^{lm\omega}(r')H(r'-r)+ R_{\rm in}^{lm\omega}(r')R_{\rm up}^{lm\omega}(r)H(r-r') \right]\,,
\eeq
satisfying ${\mathcal L}^{(r)}_{RW} G_{lm\omega}(r,r')=f(r)\delta(r-r')$, in terms of two, specially chosen, independent homogeneous solutions $R_{\rm in}^{lm\omega}(r)$ and $R_{\rm up}^{lm\omega}(r)$ of the RW operator \eqref{eq:operator}. 
Here 
\beq
W_{lm\omega}=f(r)\left[R_{\rm in}^{lm\omega}(r)R'{}_{\rm up}^{lm\omega}(r)-R'{}_{\rm in}^{lm\omega}(r)R_{\rm up}^{lm\omega}(r)\right]
\eeq
is the (constant) Wronskian and $H(x)$ denotes the Heaviside step function.
Both even-parity and odd-parity solutions
\beq
R^{\rm(even/odd)}_{lm\omega}(r)=\int d r' \, \frac{G_{lm\omega}(r,r')}{f(r')}S^{\rm(even/ odd)}_{lm\omega}(r')
\eeq
are then uniquely determined by selecting $R_{\rm in}^{lm\omega}(r)$ as the homogeneous solution which is incoming from
infinity, i.e., purely ingoing on the horizon, and $R_{\rm up}^{lm\omega}(r)$ as that one which is upgoing from the horizon, i.e., purely outgoing at infinity. 
The even source terms come with a factor $Y^*_{lm}$, while the odd ones with a factor $Y'{}^*_{lm}$, 
which can then be factored out.
Recalling that
\beq
\int h(x) \delta^{(k)}(x-x_0) dx=(-1)^k \lim_{x\to x_0} \left(\frac{d^k h(x)}{dx^k}\right)\,,
\eeq
we find, for example  (the subscript $-$ denoting a left limit $r \to r_0^{-}$) 
\beq
R^{\rm(even/odd)}_{lm\omega,-}(r)=\sum_{n,k} [s^{\rm(even/odd)}_{n,k}(r_0)\delta_n +\tilde s^{\rm(even/odd)}_{n,k}(r_0)\delta'_n] \frac{R_{\rm in}^{lm\omega}(r)}{W_{lm\omega}}(-1)^k  \lim_{r'\to r_0} \left(\frac{d^k }{dr'{}^k}\frac{R_{\rm up}^{lm\omega}(r')}{f(r')}\right)\,.
\eeq
Next, replacing the second (and higher) radial derivatives of $R_{\rm in / up}^{lm\omega}(r)$  by using the RW equation leads to expressions of the type
\begin{eqnarray}
\label{RleftvsRin}
R^{\rm(even)}_{lm\omega,-}(r) &=& \sum_n  \frac{Y_{lm}^*\left(\frac{\pi}2\right)}{W_{lm\omega}} 
 \left[J_{\rm up \, (even)}^{lm\omega, n}(r_0)\delta_n + \tilde J_{\rm up \, (even)}^{lm\omega, n}(r_0)\delta'_n\right] R_{\rm in}^{lm\omega}(r)\,,\nonumber\\
R^{\rm(odd)}_{lm\omega,-}(r) &=& \sum_n  \frac{Y'{}_{lm}^*\left(\frac{\pi}2\right)}{W_{lm\omega}} 
 \left[J_{\rm up \, (odd)}^{lm\omega, n}(r_0)\delta_n + \tilde J_{\rm up \, (odd)}^{lm\omega, n}(r_0)\delta'_n\right] R_{\rm in}^{lm\omega}(r)\,,
\end{eqnarray} 
where
\begin{eqnarray}
\label{Jupdef}
J_{\rm up \, (even/odd)}^{lm\omega, n}(r_0)&=&J_{R_{\rm up}\,{\rm(even/odd)}}^{lm\omega,n}(r_0)R^{lm\omega}_{\rm up}(r_0)+J_{R'_{\rm up}\,{\rm(even/odd)}}^{lm\omega,n}(r_0)R'{}^{lm\omega}_{\rm up}(r_0)\,,\nonumber\\
\tilde J_{\rm up \, (even/odd)}^{lm\omega, n}(r_0)&=&\tilde J_{R_{\rm up}\,{\rm(even/odd)}}^{lm\omega,n}(r_0)R^{lm\omega}_{\rm up}(r_0)+\tilde J_{R'_{\rm up}\,{\rm(even/odd)}}^{lm\omega,n}(r_0)R'{}^{lm\omega}_{\rm up}(r_0)\,,
\end{eqnarray}
and similarly for $R^{\rm(even/odd)}_{lm\omega,+}(r)$.

\subsection{Computing $h_{uk}$} 

Next, one can compute the components of the perturbed metric and, in particular, the contraction $h_{uk}$ needed to construct the Detweiler's redshift invariant.
One has at a generic spacetime point
\begin{eqnarray}
\label{eq:21}
h_{uk}(t,r,\theta,\phi) &=& \sum_{lm}[h_{uk}^{lm\,{\rm(even)}}(t,r,\theta,\phi)+h_{uk}^{lm\,{\rm(odd)}}(t,r,\theta,\phi)]\,, \nonumber\\
h_{uk}^{lm\,{\rm(even)}}(t,r,\theta,\phi)&=&h_{uk}^{lm\,{\rm(even)}}(t,r)e^{im\phi}Y_{lm}(\theta)\,, \nonumber\\
h_{uk}^{lm\,{\rm(odd)}}(t,r,\theta,\phi)&=&h_{uk}^{lm\,{\rm(odd)}}(t,r)e^{im\phi}Y'{}_{lm}(\theta)\,, 
\end{eqnarray}
and
\begin{eqnarray}
\label{eq:22}
h_{uk}^{lm\,{\rm(even)}}(t,r)&=&u^t\left[f(r)H_{0}^{lm}(t,r)+2H_{1}^{lm}(t,r)\left(\frac{dr_0}{dt}\right)+f(r)^{-1}H_{2}^{lm}(t,r)\left(\frac{dr_0}{dt}\right)^2
+r^2K^{lm}(t,r)\left(\frac{d\phi_0}{dt}\right)^2\right]\,, \nonumber\\
h_{uk}^{lm\,{\rm(odd)}}(t,r)&=&2u^t\left(\frac{d\phi_0}{dt}\right)\left[h_{0}^{lm}(t,r)+h_{1}^{lm}(t,r)\left(\frac{dr_0}{dt}\right)^2\right]\,. 
\end{eqnarray}
Each of the even/odd time-domain RWZ metric perturbations, $H_0^{lm}(t,r), \cdots$, are given by Fourier integrals, say  $H_0^{lm}(t,r) = \int d \omega  H_0^{lm \omega}(r) e^{ - i \omega t}, \cdots$, where $H_0^{lm \omega}(r), \cdots$ are linear combinations  of  the RW solutions $R_{lm\omega}^{\rm (even/odd)}(r)$ and of their first radial derivatives. 
Therefore, Fourier-transforming $h_{uk}^{lm\,{\rm(even/odd)}}(t,r)$ leads to (formally)
\beq
h_{uk\,,\pm}^{lm\omega\,{\rm(even/odd)}}(r)=K_{R_{\rm up/in}\,{\rm(even/odd)}}^{lm\omega}(r)\, R_{lm\omega\,,\pm}^{\rm(even/odd)}(r)
+K_{R'{}_{\rm up/in}\,{\rm(even/odd)}}^{lm\omega}(r)\,  R'{}^{\rm(even/odd)}_{lm\omega\,,\pm}(r)\,.
\eeq

Consider, for instance, the even contribution to the left part, i.e., $h_{uk\,,-}^{lm\omega\,{\rm(even)}}(r)$.
First replace $R^{\rm(even)}_{lm\omega,-}$ with $R_{\rm in}^{lm\omega}$ through Eqs. \eqref{RleftvsRin}, \eqref{Jupdef} to obtain
\begin{eqnarray}
h_{uk\,,-}^{lm\omega\,{\rm(even)}}(r)&=&\sum_n  \frac{Y_{lm}^*\left(\frac{\pi}2\right)}{W_{lm\omega}} 
 \left[J_{\rm up \, (even)}^{lm\omega, n}(r_0)\delta_n + \tilde J_{\rm up \, (even)}^{lm\omega, n}(r_0)\delta'_n\right]
\left[K_{R_{\rm in}\,{\rm(even)}}^{lm\omega}(r)\, R^{lm\omega}_{\rm in}(r)
+K_{R'_{\rm in}\,{\rm(even)}}^{lm\omega}(r)\,  R'{}^{lm\omega}_{\rm in}(r)\right]\nonumber\\
&\equiv&
\sum_n  Y_{lm}^*\left(\frac{\pi}2\right)\left\{  \left[C_1^{lm\omega, n}(r)\delta_n + \tilde C_1^{lm\omega, n}(r)\delta'_n\right]Z_1^{lm\omega}(r) +\ldots \right\}\,,
\end{eqnarray}
where
\begin{eqnarray}
&&C_1^{lm\omega, n}(r)=J_{R_{\rm up}\,{\rm (even)}}^{lm\omega,n}(r_0)K_{R_{\rm in}\,{\rm (even)}}^{lm\omega}(r)\,,\quad
C_2^{lm\omega, n}(r)=J_{R'_{\rm up}\,{\rm (even)}}^{lm\omega,n}(r_0)K_{R_{\rm in}\,{\rm (even)}}^{lm\omega}(r)\,,\nonumber\\
&&C_3^{lm\omega, n}(r)=J_{R_{\rm up}\,{\rm (even)}}^{lm\omega,n}(r_0)K_{R'_{\rm in}\,{\rm (even)}}^{lm\omega}(r)\,,\quad
C_4^{lm\omega, n}(r)=J_{R'_{\rm up}\,{\rm (even)}}^{lm\omega,n}(r_0)K_{R'_{\rm in}\,{\rm (even)}}^{lm\omega}(r)\,,\nonumber\\
&&\tilde C_1^{lm\omega, n}(r)=\tilde J_{R_{\rm up}\,{\rm (even)}}^{lm\omega,n}(r_0)K_{R_{\rm in}\,{\rm (even)}}^{lm\omega}(r)\,,\quad
\tilde C_2^{lm\omega, n}(r)=\tilde J_{R'_{\rm up}\,{\rm (even)}}^{lm\omega,n}(r_0)K_{R_{\rm in}\,{\rm (even)}}^{lm\omega}(r)\,,\nonumber\\
&&\tilde C_3^{lm\omega, n}(r)=\tilde J_{R_{\rm up}\,{\rm (even)}}^{lm\omega,n}(r_0)K_{R'_{\rm in}\,{\rm (even)}}^{lm\omega}(r)\,,\quad
\tilde C_4^{lm\omega, n}(r)=\tilde J_{R'_{\rm up}\,{\rm (even)}}^{lm\omega,n}(r_0)K_{R'_{\rm in}\,{\rm (even)}}^{lm\omega}(r)\,,\nonumber\\
&&Z_1^{lm\omega}(r)=R^{lm\omega}_{\rm in}(r)R^{lm\omega}_{\rm up}(r_0)W_{lm\omega}^{-1}\,,\quad
Z_2^{lm\omega}(r)=R^{lm\omega}_{\rm in}(r)R'{}^{lm\omega}_{\rm up}(r_0)W_{lm\omega}^{-1}\,,\nonumber\\
&&Z_3^{lm\omega}(r)=R'{}^{lm\omega}_{\rm in}(r)R^{lm\omega}_{\rm up}(r_0)W_{lm\omega}^{-1}\,,\quad
Z_4^{lm\omega}(r)=R'{}^{lm\omega}_{\rm in}(r)R'{}^{lm\omega}_{\rm up}(r_0)W_{lm\omega}^{-1}\,.
\end{eqnarray}
Properly transforming terms according to 
\beq
f(\omega)\delta_n=f(\omega_{m,n})\delta_n\,,\qquad
f(\omega)\delta'_n=-f'(\omega_{m,n})\delta_n+f(\omega_{m,n})\delta'_n\,,
\eeq
and Fourier anti-transforming 
\beq
h_{uk\,,-}^{lm\,{\rm(even)}}(t,r)=\frac1{2\pi}\int e^{-i\omega t}h_{uk\,,-}^{lm\omega\,{\rm(even)}}(r)d\omega\,,
\eeq
then yields
\begin{eqnarray}
\label{huuleftevensol}
h_{uk\,,-}^{\rm(even)}(t)&=&
\frac{1}{2\pi}\sum_{lmn}  \left|Y_{lm}\left(\frac{\pi}2\right)\right|^2e^{i(m\phi_0(t)-\omega_{m,n} t)}\left\{ 
\left[\left(C_1^{lm, n} + it\tilde C_1^{lm, n} - \frac{d\tilde C_1^{lm\omega, n}}{d\omega}\bigg\vert_{\omega=\omega_{m,n}}\right)Z_1^{lm,n}\right.\right.\nonumber\\
&&\left.\left.
-\tilde C_1^{lm, n}\frac{dZ_1^{lm\omega}}{d\omega}\bigg\vert_{\omega=\omega_{m,n}}\right]_{r=r_0(t)}
+\ldots
\right\}\,,
\end{eqnarray}
where all quantities are evaluated at the particle position $r=r_0(t)$.
A similar procedure applies to the odd case.

The resulting $h_{uk}(t)$ has to be averaged over a (coordinate-time) radial period, and once inserted in Eq. \eqref{delta_U1} finally gives $\delta U$. The latter should be suitably regularized in order to remove its singular part.
Barack and Sago \cite{Barack:2011ed} provided a closed form expression (in terms of elliptic integrals) for the regularization parameter $B_H$ of the quantity $H=\frac12 h_{uu}$ (see their Eq. (D15)), which is still a function of time, being evaluated at the particle position.
A feature of our formalism is that, in order to compute the {\it regularized} value of $\langle h_{uk}\rangle_t$, we do not need
to analytically determine in advance the corresponding subtraction term, because we automatically obtain it as a side-product of our computation [by taking the 
$l\to \infty$ limit of our PN-based calculation.]
The subtraction term (a.k.a. ``regularization parameter") for the quantity $U_0\langle h_{uk}\rangle_t=\langle h_{uu}\rangle_\tau$ is 
\begin{eqnarray}
B&=&  2 u_p-\frac12 u_p^2-\frac{39}{32}u_p^3-\frac{385}{128}u_p^4-\frac{61559}{8192}u_p^5-\frac{622545}{32768}u_p^6-\frac{25472511}{524288}u_p^7-\frac{263402721}{2097152}u_p^8-\frac{176103411255}{536870912}u_p^9 \nonumber\\
&&+  \left(-2 u_p+\frac74 u_p^2+7 u_p^3+\frac{8597}{256}u_p^4+\frac{1498513}{8192}u_p^5+\frac{69481763}{65536}u_p^6+\frac{1650414477}{262144} u_p^7+\frac{158088550401}{4194304}u_p^8\right.\nonumber\\
&&\left.
 +\frac{121418022556683}{536870912}u_p^9\right)e^2\nonumber\\
&& +\left( -\frac{23}{16} u_p^2-\frac{1655}{256}u_p^3-\frac{16549}{512} u_p^4-\frac{5554769}{32768} u_p^5-\frac{229907593}{262144} u_p^6-\frac{17904332713}{4194304} u_p^7-\frac{77183281089}{4194304} u_p^8\right.\nonumber\\
&&\left.
-\frac{63794507176773}{1073741824} u_p^9 \right)e^4
 +O(u_p^{10},e^6)\,.
\end{eqnarray}
It is simply related to that for $\delta U$ by $B_{\delta U}=\frac12 U_0 B$.
We checked that our (PN- and eccentricity-expanded) result for $B$ agrees with the correspondingly-expanded 
(time-averaged) analytical result derived in \cite{Barack:2011ed}, via $B=2\langle B_H\rangle_\tau$.

The low multipoles ($l=0,1$) have been computed separately (using the method of \cite{Keidl:2010pm}). Their (already regularized) contribution to $\delta U$ is the following

\begin{eqnarray}
\delta U^{l=0,1}&=&- u_p-\frac12 u_p^2-\frac{57}{32} u_p^3-\frac{1325}{128} u_p^4-\frac{417289}{8192} u_p^5-\frac{7203141}{32768} u_p^6-\frac{457801857}{524288} u_p^7 -\frac{6888106557}{2097152} u_p^8\nonumber\\
&&
-\frac{6386307327945}{536870912} u_p^9\nonumber\\
&+&\left(u_p+\frac94 u_p^2+\frac78 u_p^3-\frac{7473}{256} u_p^4-\frac{2080289}{8192} u_p^5-\frac{108723171}{65536} u_p^6-\frac{2615543903}{262144} u_p^7-\frac{245438383189}{4194304} u_p^8\right.\nonumber\\
&&\left.
-\frac{183533858669131}{536870912}u_p^9\right) e^2\nonumber\\
&+& \left(-\frac{33}{16} u_p^2+\frac{895}{256} u_p^3+\frac{18815}{256} u_p^4+\frac{16321729}{32768} u_p^5+\frac{678595077}{262144} u_p^6+\frac{49395616017}{4194304} u_p^7+\frac{401066889193}{8388608} u_p^8\right.\nonumber\\
&&\left.
+\frac{171799575733669}{1073741824} u_p^9\right) e^4
+O(u_p^{10},e^6)\,.
\end{eqnarray}

\section{Coefficients of the PN expansions of the functions $a(u_3)$, $\bar d(u_3)$, $\rho(u_3)$, and $q(u_3)$
(where $u_3 \equiv 3 u$).}
\label{appTables}

We list  below in Tables \ref{hata}, \ref{hatrho}, \ref{hatbard} and \ref{hatq} the coefficients of the PN expansions of the functions $a(u_3)$, $\rho(u_3)$, $\bar d(u_3)$ and $q(u_3)$, respectively, where $u_3 \equiv 3 u$.
For instance, the PN expansion of the EOB radial potential $a(u)$ in powers of $u_3$ is defined by Eq. \eqref{au3pnexp}, and similarly for the other EOB functions.

\begin{table}[h]
\centering
\caption{Coefficients of the PN expansion of $a(u_3)$ (where $u_3 \equiv 3 u$).}
\begin{ruledtabular}
\begin{tabular}{lllllll}
$N$ & ${ \widehat  a}_N $ & ${\widehat  a}_N'$  & $ {\widehat  a}_N''$ & ${\widehat  a}_N'''$& ${\widehat  a}_N''''$ \cr
\hline
3& 0.7407407407$\times 10^{-1}$& 0& 0& 0& 0\cr 
4& 0.2307148148& 0& 0& 0& 0\cr 
5& 0.3885272716$\times 10^{-1}$& 0.5267489711$\times 10^{-1}$& 0& 0& 0\cr 
6& -0.8338706215$\times 10^{-1}$& -0.9150172836$\times 10^{-1}$& 0& 0& 0\cr 
7& 0.3407031724& -0.2847361682$\times 10^{-2}$& 0& 0& 0\cr 
8& 0.1039436274& 0.1598400803& -0.7952324342$\times 10^{-2}$& 0& 0\cr 
9& -0.4718506679& -0.8953034274$\times 10^{-1}$& 0.1255096276$\times 10^{-1}$& 0& 0\cr 
10& 0.2384172903& -0.1456465064& 0.3027617741$\times 10^{-2}$& 0& 0\cr 
11& 0.6581815598& 0.1243999564& -0.2520471365$\times 10^{-1}$& 0.8003748299$\times 10^{-3}$& 0\cr 
12& -0.4387682766& 0.1582428946& 0.8613195585$\times 10^{-2}$& -0.1083217892$\times 10^{-2}$& 0\cr 
13& -0.4889356138& -0.1260270367& 0.2542670993$\times 10^{-1}$& -0.6009925216$\times 10^{-3}$& 0\cr 
14& 0.5006006303& -0.1885956666& -0.1182152359$\times 10^{-1}$& 0.2474966882$\times 10^{-2}$& -0.6041644844$\times 10^{-4}$\cr 
15& 0.5018616919& 0.1112996214& -0.2899966966$\times 10^{-1}$& -0.706112373$\times 10^{-4}$& 0.6307853274$\times 10^{-4}$\cr 
16& -0.1446511265& 0.2496830421& 0.995357387$\times 10^{-3}$& -0.2455369950$\times 10^{-2}$& 0.6726163302$\times 10^{-4}$\cr 
17& -0.6989207457& -0.2953308384$\times 10^{-1}$& 0.3736873206$\times 10^{-1}$& -0.3176428511$\times 10^{-3}$& -0.1602661295$\times 10^{-3}$\cr 
18& -0.2814052401& -0.3589800307& 0.1913120432$\times 10^{-1}$& 0.2652093198$\times 10^{-2}$& -0.7390733724$\times 10^{-4}$\cr 
19& 1.382632670& -0.1428658618& -0.5474406082$\times 10^{-1}$& 0.3892284877$\times 10^{-2}$& 0.8759119256$\times 10^{-4}$\cr 
20& 0.8449305382& 0.5415180408& -0.6363323976$\times 10^{-1}$& -0.2718419238$\times 10^{-2}$& 0.1900895693$\times 10^{-3}$\cr 
21& -2.254005320& 0.4263215985& 0.7445208813$\times 10^{-1}$& -0.9823954351$\times 10^{-2}$& 0.3111490819$\times 10^{-4}$\cr 
22& -1.459428785& -0.8106346246& 0.1510854162& 0.2828816180$\times 10^{-2}$& -0.8152502117$\times 10^{-3}$\cr 
23& 3.312652006& -0.9193948736& -0.9370003950$\times 10^{-1}$& 0.2168105861$\times 10^{-1}$& -0.4310759893$\times 10^{-3}$\cr
\end{tabular}
\end{ruledtabular}
\label{hata}
\end{table}

\begin{table}[h]
\centering
\caption{$(N-1)^2$-rescaled coefficients of the PN expansion of $\rho(u_3)$.}
\begin{ruledtabular}
\begin{tabular}{llll}
$N$ & ${\widehat \rho}_N/(N-1)^2$ & ${\widehat \rho}_N'/(N-1)^2$  & ${\widehat \rho}_N''/(N-1)^2$   \cr
\hline
2& 1.555555556& 0& 0\cr 
3& 1.135439039& 0& 0\cr               
4& -0.163704216&  0.229721079 & 0\cr 
5& -0.735645815& -0.416519694 & 0\cr 
6& 1.349760034&  0.104713538 & 0\cr 
7& -0.126387674&  0.583029222 & -0.273913594$\times 10^{-1}$\cr 
8&-2.325715312& -0.676167975 &  0.646912737$\times 10^{-1}$\cr 
9& 2.513342107& -0.231444902 & -0.312270551$\times 10^{-1}$\cr
\end{tabular}
\end{ruledtabular}
\label{hatrho}
\end{table}

\begin{table}[h]
\centering
\caption{$(N-1)^2$-rescaled coefficients of the PN expansion of $\bar d(u_3)$.}
\begin{ruledtabular}
\begin{tabular}{llll}
$N$ & $\widehat {\bar d}_N/(N-1)^2$ & $\widehat {\bar d}_N'/(N-1)^2$  & $\widehat {\bar d}_N''/(N-1)^2$   \cr
\hline
2& 0.666666667& 0& 0\cr 
3& 0.481481481& 0& 0\cr 
4& 0.244462862                 & 0.541380887$\times 10^{-1}$& 0\cr 
5& -0.6597928$\times 10^{-1}$ &-0.521751911$\times 10^{-1}$& 0\cr 
6& 0.299039195                &-0.390769700$\times 10^{-1}$& 0\cr 
7& 0.270278611&                 0.163046626& -0.883592239$\times 10^{-2}$ \cr 
8&-0.696093391&                  -0.103547802 & 0.164167717 $\times 10^{-1}$\cr 
9& 0.531333085&                  -0.165934019&               -0.316508395$\times 10^{-3}$\cr  
\end{tabular}
\end{ruledtabular}
\label{hatbard}
\end{table}

\begin{table}[h]
\centering
\caption{$(N-1)^4$-rescaled coefficients of the PN expansion of $q(u_3)$.}
\begin{ruledtabular}
\begin{tabular}{llll}
$N$ & ${\widehat q}_N/(N-1)^4$ & $\widehat q_N'/(N-1)^4$  & $\widehat {q}_N''/(N-1)^4$   \cr
\hline
2&  0.888888889& 0& 0\cr 
3& 0.664607507$\times 10^{-1}$& 0& 0\cr 
4&-0.234627665$\times 10^{-1}$&  0.787917057$\times 10^{-2}$& 0\cr 
5& 0.155706161$\times 10^{-1}$& -0.157616639$\times 10^{-1}$& 0\cr 
6& 0.683824536$\times 10^{-1}$&  0.327993634$\times 10^{-1}$& -0.141710523$\times 10^{-2}$\cr 
7& -0.221679525&    -0.332538735$\times 10^{-1}$&  0.405402116$\times 10^{-2}$\cr 
8&  0.262385007&    -0.296664893$\times 10^{-1}$& -0.274350852$\times 10^{-2}$\cr
\end{tabular}
\end{ruledtabular}
\label{hatq}
\end{table}

\end{widetext}

\subsection*{Acknowledgments}
We thank Maarten van de Meent for informative email exchanges  about the precession function $\rho(u)$.
D.B. thanks the Italian INFN (Naples) for partial support and IHES for hospitality during the development of this project.
All  the authors are grateful to ICRANet for partial support.

\end{document}